\documentclass[fleqn,10pt,twocolumn]{wlscirep}
\usepackage[utf8]{inputenc}
\usepackage[T1]{fontenc}
\usepackage{xr}
\usepackage{hyperref}
\usepackage{tabularx}
\usepackage{cuted}
\usepackage[numbers,sort&compress]{natbib}
\usepackage{graphicx}
\usepackage{float}
\usepackage{dcolumn}
\usepackage{bm}
\usepackage{soul}
\usepackage{amsmath}
\usepackage{amssymb}
\usepackage{color}
\usepackage{multirow}
\usepackage{setspace}
\usepackage{array}
\usepackage{booktabs}
\usepackage{rotating}
\usepackage{ulem}
\usepackage{physics}
\usepackage{mathrsfs}
\usepackage{amssymb,amsfonts,dsfont}
\usepackage{bbold}
\usepackage{times} 
\usepackage{graphicx} 
\usepackage{accents} 
\usepackage{epic}
\usepackage{wrapfig}
\usepackage[capitalise]{cleveref}
\usepackage[english]{babel}

\definecolor{Grey}{rgb}{0.50,0.50,0.50}
\definecolor{Blu}{rgb}{0.00,0.00,1.00}
\definecolor{Red}{rgb}{1.00,0.00,0.00}
\definecolor{Green}{rgb}{0.20,0.50,0.20}
\definecolor{Magenta}{rgb}{0.60,0.00,0.60}
\definecolor{BluBondi}{rgb}{0.00,0.58,0.71}
\definecolor{Orange}{rgb}{0.95,0.46,0.17}
\definecolor{Red}{rgb}{1.00,0.00,0.00}

\newcommand{\editor}[2]{%
  \expandafter\newcommand\csname #1note\endcsname[1]{%
    \textcolor{#2}{(\textbf{#1:} \textit{##1})}}%
  \expandafter\newcommand\csname #1\endcsname[1]{%
    \textcolor{#2}{##1}}%
  \expandafter\newcommand\csname #1cancel\endcsname[1]{%
    \textcolor{#2}{\sout{##1}}}%
  \expandafter\newcommand\csname #1change\endcsname[2]{%
    \textcolor{#2}{\sout{##1} ##2}}%
  \newenvironment{#1text}{\color{#2}}{\color{black}}
}

\editor{PDA}{cyan}
\editor{GS}{Red}
\editor{AG}{Green}
\editor{AF}{red}
\editor{DV}{Orange}
\editor{MR}{Magenta}
\editor{CC}{BluBondi}

\newcommand{\drop}[1]{}

\newcommand{\kv}{\boldsymbol{k}}

\newcommand{\qv}{\boldsymbol{q}}

\newcommand{\yambo}{\textsc{yambo}}
\newcommand{\yambopy}{\textsc{yambopy}}
\title{A viable and accurate route to GW simulations of metals and doped semiconductors}

\author[1,2]{Giacomo Sesti}
\affil[*]{giacomo.sesti@nano.cnr.it}

\author[3]{Alberto Guandalini}
\author[1]{Andrea Ferretti}
\author[1]{Pino D'Amico}
\author[1]{Claudia Cardoso}
\affil[**]{claudia.cardoso@nano.cnr.it}
\author[1]{Massimo Rontani}
\author[1]{Daniele Varsano}

\affil[1]{Istituto Nanoscienze- CNR, S3, Via G. Campi 213/A, Modena (Italy)}
\affil[2]{FIM Department, Universit\`a degli Studi di Modena e Reggio Emilia, Via Campi 213a, 41125 Modena, Italy}
\affil[3]{Dipartimento di Fisica, Universit\`a di Roma La Sapienza, Piazzale Aldo Moro 5, I-00185 Roma, Italy}

\begin{abstract}
In this work we present an efficient and fully ab-initio approach for the calculation of the screened Coulomb potential, specifically designed for two dimensional (2D) and three dimensional (3D) metals and suitable to evaluate $GW$ corrections. While quasiparticle renormalization effects are typically negligible in bulk metals, 2D metals may exhibit semiconductor-like features, such as sizeable renormalization of quasiparticle energies induced by Coulomb interaction, which follow from the suppression of screening due to the reduced dimensionality, Although this effect can in principle be described by the $GW$ method, apart from a few pioneering exceptions, such calculations have been hampered by sampling requirements exceeding feasibility. We demonstrate that the combination of interpolation and extrapolation schemes together with the accounting of  dynamical effects is pivotal to the correct handling of the intraband polarizability in the long wavelength limit. As an application, we report the gap renormalization of selected doped 2D-semiconductors in excellent agreement with ARPES measurements.
\end{abstract}

\begin{document}
\maketitle

\section{Introduction}

In two dimensions (2D), the screening of the Coulomb interaction among charge carriers is heavily suppressed with respect to the bulk (3D), since  the electric field lines spill over the vacuum. As a consequence, even metals may exhibit signatures of strong electron correlations, leading to exotic phases such as unconventional superconductors~\cite{Little1964,Ginzburg1968,Bardeen1973,Sham1983,Sham1984,Littlewood1990,vanWezel2010,Laussy2010,Ilani2016,Crepel2022,jindal_coupled_2023,Wu2024} and ferroelectrics~\cite{Portengen1996b,fei_ferroelectric_2018,Varsano2020,Ataei2021,zheng_2023,DAlessio2025},
or 
excitonic insulators~\cite{DiSalvo1976,Khveshchenko2001,Cercellier2007,Kogar2017,Kono2017,Ma2021,sun_evidence_2022,jia_evidence_2022,Gao2023,Gao2024,Hossain2025,Qi2025}.
In this work we focus on the renormalization of quasiparticle energies as driven by electronic correlations,
an effect that is usually associated with semiconductors and originates from the long range of the poorly screened Coulomb forces.

In this respect, many-body perturbation theory is the method of choice to provide accurate, first principles predictions of the quasiparticle band structure~\cite{Onida2002RMP,martin2016book,Marzari2021NatureMat}, 
within the the so-called $GW$ approximation~\cite{Hedin1965PR, Strinati_1982, Aryasetiawan1998RPP, martin2016book, Reining2018wcms, Golze2019FrontChem}.
Whereas this method
proved to be extremely successful for semiconductors~\cite{Kohn1965PhysRev,Marzari2021NatureMat, Reining2018wcms, Golze2019FrontChem, Louie2021NatMater, vanSchilfgaarde2006PhysRevLett, Blase2011PhysRevB}, its feasibility for metals is hampered by $\kv$-point and frequency sampling requirements which are typically out of reach, with a few remarkable exceptions~\cite{Spataru2004, Yang2007, Spataru2010, Spataru2013, DAmico2020PRB}. Even standard textbook results,
like the softening of the 2D plasmon at long wavelengths, are reproduced only through
{\it ad hoc} adjustments of the computational protocol~\cite{Liang2015, Gao2017} or via a significant increase of the $\kv$-grid density~\cite{da2020universal}.

In this work, we report a viable and fully \textit{ab initio} $GW$ methodology for quasi-particle calculations in the presence of a Fermi surface, which includes both metals and doped semiconductors in 2D and 3D. Our unified implementation, based on a new development within the stochastic W-average (W-av) method~\cite{Guandalini2023npjCM,GuandaliniLeon_2024}, accurately includes the intraband contributions to the response function. Importantly, our approach eliminates the need for dense $\kv$-grids, and remains fully compatible with the use of an accurate frequency description of the screened potential. The proposed method is applied in a non-self consistent framework ($G_0W_0$), but the method can  be straightforwardly extended to other schemes, such as self-consistent $GW$. As a validation, we report excellent agreement of the doping-dependent band gap renormalization of monolayer MoS$_2$, as compared to photoemission spectra \cite{LiuMoS2}.

We emphasize that the proposed $GW$ methodology copes with all computational challenges of doped semiconductors and metals, as follows.
For semiconductors, doping can induce surprisingly large band-gaps renormalization. Critically, the strong coupling between the doping-induced carrier plasmon and the quasiparticle excitations requires extra care in the calculation of the many-electron screening that determines the dielectric function.
In order to describe the intraband plasmon associated with the free carriers,
it is crucial to go beyond single pole frequency descriptions such as the plasmon pole approximation (PPA)~\cite{Hybertsen1986PRB,vonderLinden1988PRB,Northrup1989PRB, Godby1989PRL,Engel1993PRB}.
Previous works have resorted to the inclusion of an additional pole when modeling the dielectric function~\cite{Liang2015, Gao2017, Champagne2023}. 
Additionally, for bulk metallic systems~\cite{mandal_2022, friedrich_2022}, the excited-state spectra obtained from angle-resolved photoemission spectroscopy (ARPES) are frequently significantly narrower than those predicted by Kohn-Sham density functional theory (DFT) with local or semilocal approximations~\cite{Lyo_88}. The $GW$ approximation has been shown to improve DFT results also in this case, provided that the methodological challenges in the description of metallic screening are properly addressed~\cite{Wooten1972book, Marini2001PRB,Cazzaniga2012PRB,Leon_2023}.

In metals, electronic excitations within the same band lead to poles in the irreducible polarizability $\chi_0$, whose energies vanish for $\qv \rightarrow 0$, while the Coulomb potential $v$ diverges on the same scale. The screening depends on the integration of the product of these two quantities on all metallic bands ~\cite{Marini2001PRL, Krystyna_2020}, which can be challenging since the discretization of the Brillouin zone (BZ) can lead to ill-defined Fermi surfaces.
These contributions from the Fermi surface are  typically neglected, and their impact in the screening is minimized by increasing the $\kv$-point density sampling of the BZ, which can significantly slow down convergence~\cite{Methfessel1989PRB,Maksimov1988JPEMP,Lee1994PRB,Marini2001PRB}, causing, for example, spurious band gaps that vanish very slowly with increasing number of $\kv$-points~\cite{Cazzaniga2008PRB}. 
Analytical models using a Taylor expansion of the dielectric function for small-$\qv$, require fewer $\kv$-points~\cite{Cazzaniga2008PRB,Cazzaniga2010PRB,Orhan2019JPCM}, but often depend on external parameters.
This is the case of Drude-based approaches~\cite{Lee1994PRB}, where the missing intra-band contribution is accounted for by adding a phenomenological term to the head of the irreducible dielectric matrix in the $\qv\to0$ limit. The Drude term depends on a plasma frequency $\omega_D$ and a damping factor $\gamma$, parameters that can be determined either experimentally or by expensive {\it ab-initio} methods~\cite{Lee1994PRB, Marini2001PRL, Kohn1974PRB, Sporkmann1994PRB, Prandini2019ComputPhysCommun, Prandini2019npjComputMater, Methfessel1989PRB, Blochl1994PRB, Lee1994PRB, Friedrichnano2022}.  

For bulk metals, the dielectric function at $\qv=0$ has a slow quadratic dispersion, which allows the treatment of intra-band transitions with the so-called constant approximation, introduced by our group~\cite{Leon_2023}.
In contrast, for 2D systems, the dielectric function in the long-wavelength limit exhibits large variations with $\qv$, requiring very dense $\kv$-grids to accurately integrate over the BZ~\cite{qiu2016screening, Huser_2013}. The picture results then similar to the case of 2D semiconductors for which several schemes have been proposed to mitigate this problem~\cite{Rasmussen_2016,daJornada_2017,Xia_2020}, including our previous work where we introduced the W-av scheme~\cite{Guandalini2023npjCM}. In that work we showed that, combining a Monte Carlo integration with an appropriate interpolation of the screened potential between the calculated grid points drastically accelerates convergence with respect to the BZ sampling. However, the case of metals was not addressed at that point.

In this work, we address the problems described above in the calculation of $G_0W_0$ quasiparticle (QP) corrections for systems with metallic screening, extending the W-average methodology to provide a consistent and accurate treatment of both 2D and 3D materials, including doped semiconductors.  
The paper is organized as follows. In Section~\nameref{s:primer} we open by summarizing the main steps of the $GW$ approximation and describe the W-av approach, as previously developed for semiconductors. In Section~\nameref{methods:Wav_metals}, we present the new development within the W-av method, now compliant with the description of metallic screening. We then discuss results enabled by the methodology on selected systems. First we discuss the case of two doped monolayers, MoS$_2$ and graphene, including results concerning band gap renormalization in 2D (see Section~\nameref{section:bandgap_renorm_2D}) and then we validate the method for two bulk metals, Na and Mg (see Section~\nameref{section:results_3d}). 
Section~\nameref{sec:3} provides more details on the W-av procedure and the accuracy of the method. Finally,
in the~\nameref{section:conclusions} we draw the main conclusions, and in the~\nameref{sec:methods} we summarize the computational parameters used in the calculations.

\section{Results and discussions}
\label{section:results}
%
\subsection{Theoretical framework}\label{s:primer}

Within Green's function methods from many-body perturbation theory, the quasi-particle energies $E_{n\kv}$ 
can be calculated perturbatively from the Kohn-Sham (KS) single-particle energies $\varepsilon_{n\kv}$ by replacing the exchange correlation potential $V_{xc}$ with the diagonal elements of the self-energy $\Sigma$ evaluated on the Bloch states $| n\kv \rangle$:
\begin{equation}
    \label{eq.gw_ex}
    E_{n\kv}= \varepsilon_{n\kv}+ \langle n\kv | \Sigma( E_{n\kv})-V_{xc}-\Delta \Sigma| n\kv \rangle,  
\end{equation}
where $\Delta \Sigma=\Sigma(\mu)-V_{xc}(\mu)$ is the quasi-particle correction at the chemical potential $\mu$~ \cite{Hedin1999}. When dealing with metals, one must use $\mu$ as a reference since it aligns the quasi-particle and the single-particle Fermi levels and, for theories that conserve the particle-number, leads to a trivially zero $\Delta \Sigma$. The non-linear equation~(\ref{eq.gw_ex}) is often approximated by its linear expansion,
\begin{equation}
    \label{eq.lin}
    E_{n\kv}= \varepsilon_{n\kv}+ Z_{n\kv} \langle n\kv | \Sigma(\varepsilon_{n\kv})-V_{xc}-\Delta \Sigma | n\kv \rangle,  
\end{equation}
where $Z_{n\kv}$ is the renormalization factor
\begin{equation}
Z_{n\kv}=\left(1-\left. \frac{d \Sigma(\omega)}{d \omega} \right|_{\omega=\varepsilon_{n\kv}}\right)^{-1}.
\end{equation} 
Within the present work, for numerical stability, we consider the Taylor expansion of Re$[\Sigma]$.
The linear expansion applies if Re$[\Sigma]$ is approximately linear around $E_{nk}$, which corresponds to values of $Z_{n\kv}$ typically larger than 0.5-0.6.
Alternatively, QP energies can be determined from the position of the peaks in the spectral function $A(\omega)$, defined as the imaginary part of $G$ (expressed here neglecting the off-diagonal matrix elements of the self-energy):
\begin{equation}
\label{eq.spectralfunc}
    A_{n\kv}(\omega) = \frac{1}{\pi}\left| \text{Im} \frac{1}{\omega-\varepsilon_{n\kv}-\Sigma_{n\kv}(\omega)+V_{xc}^{(n \kv)}+\Delta \Sigma} \right|.
\end{equation}
A widely used method to calculate the self-energy is the non-iterative G$_0$W$_0$ approximation. Taking advantage of the Lehmann representation for the non-interacting Green function $G_0$ in the Bloch plane-wave basis set, the diagonal matrix elements of $\Sigma$ can be written through a double integral in frequency $\omega'$ and momentum $\qv$:
\begin{multline}
\label{eq_GW_expl}
   \Sigma_{n\kv}(\omega)  = 
   -\sum\limits_{\mathbf{G}\mathbf{G}'}
    \int\frac{d\omega'}{2 \pi i} e^{i \omega' \eta} 
     \int\frac{d\qv}{(2\pi)^3} 
     \\ \ \quad \times \,\, g^{nk}_{\mathbf{G}\mathbf{G}'}(\qv,\omega,\omega') \, W_{\mathbf{G}\mathbf{G}'}(\qv,\omega'),
\end{multline}
where 
\begin{equation}
     g^{nk}_{\mathbf{G}\mathbf{G}'}(\qv,\omega,\omega') = \sum\limits_m \frac{\rho_{nm}(\kv,\qv,\mathbf{G}) \rho^*_{nm}(\kv,\qv,\mathbf{G}')}{\omega+\omega'-\varepsilon_{m\kv-\qv}+i\eta \  \textrm{sgn}(\varepsilon_{m\kv-\qv}-\mu)}.
\end{equation}
The factors $g^{nk}_{\mathbf{G}\mathbf{G}'}(\qv,\omega,\omega')$ gather the dependencies on the plane waves and frequency originating from $G_0$. In the expressions above, $\rho_{nm}(\kv,\qv,\mathbf{G})=  \langle{n\kv}|e^{i(\qv+\mathbf{G}) \cdot \mathbf{r}}| m\mathbf{k-q} \rangle $ are the generalized oscillator strengths and $W_{\mathbf{G}\mathbf{G}'}(\qv,\omega)$ is the screened Coulomb potential. The index $m$ runs over all bands, while $\mathbf{G}$ and $\mathbf{G'}$ are plane-waves used to represent the spatial degrees of freedom. The time order is ensured by the implicit limit $\eta \to 0^+$. 

The frequency dependence of $W$ is given by the density-density response function $\chi$:
\begin{equation}
\label{eq.W}
    W_{\mathbf{G}\mathbf{G'}}(\qv,\omega) = \big[\delta_{\mathbf{G}\mathbf{G'}}+ V_{\mathbf{G}}(\qv)  \, \chi_{\mathbf{G}\mathbf{G'}}(\qv,\omega) \big]\, V_{\mathbf{G'}}(\qv),
\end{equation}
where $V_{\mathbf{G}}$ is the Fourier transform of the Coulomb potential, possibly truncated~\cite{Rozzi2006PRB,ismail2006truncation} when dealing with low-dimensional system. Here, the polarization matrix $\chi$ is computed in the random phase approximation (RPA). In this framework, the frequency dependence of $\chi$ can be determined through different methods, each offering a particular balance between accuracy and computational cost. In this work, we adopt the multipole approximation (MPA)~\cite{Leon2021PRB,Leon_2023,GuandaliniLeon_2024}, which yields results comparable to full-frequency (FF) approaches while being significantly more computationally efficient. In the MPA method, the matrix elements of $\chi(\omega)$ are expressed as the sum of a small number of complex poles: \begin{equation}
\tag{8}
    \chi_{\mathbf{G}\mathbf{G'}}(\qv,\omega)=\sum_{p=1}^{n_p}\frac{2 R_{p\mathbf{G}\mathbf{G'}}(\qv)\Omega_{p\mathbf{G}\mathbf{G'}}(\qv)}{\omega^2-[\Omega_{p\mathbf{G}\mathbf{G'}}(\qv)]^2},
\end{equation}
where $\Omega_{p\mathbf{G}\mathbf{G'}}(\qv)$ and $R_{p\mathbf{G}\mathbf{G'}}(\qv)$ are the poles and residues, respectively, and $n_p$ is the total number of poles. The residues and poles are determined for each momentum transfer $\qv$ and each element $\mathbf{G}\mathbf{G'}$ through a nonlinear interpolation with frequency points sampled in the complex plane, $z=\omega+i\bar{\omega}$, as described in detail in~\cite{Leon2021PRB}. 
The MPA effectively generalizes the PPA solution to a multipole representation of $\chi(\omega)$. Conceptually, the MPA lies between the PPA, recovered in the one-pole limit, and the exact full-frequency (FF) approach, if the number of poles increases. Notably, while the f-sum rule is not explicitly imposed, one can expect it to be recovered by MPA when reaching convergence, as e.g. exploited in Ref.~\cite{Leon_2023}, besides deviations due to non-local pseudo-potentials~\cite{Engel1992PRB}.

Following from the decomposition of the screened Coulomb potential in the exchange and correlation terms, $W(\omega) = V + W^{c}(\omega)$, the self-energy in Eq.~\eqref{eq_GW_expl} is commonly written as the sum of a static part $\Sigma^x_{n\kv}$, which depends on the bare Coulomb potential $V$, and of a dynamic part $\Sigma^c_{n\kv}(\omega)$, depending on $W^c(\omega)$. 
Within the W-av method~\cite{Guandalini2023npjCM}, the integration of $\Sigma^c_{n\kv}(\omega)$ over the momentum $\qv$ and the frequency $\omega$
can be performed separately, since $g^{nk}_{\mathbf{G}\mathbf{G}'}(\qv,\omega,\omega')$ are well-defined and smooth with respect to $\qv$. The integral in momentum follows the Monkhorst-Pack scheme~\cite{monkhorst1976special}, i.e., $\qv$ is discretised on a uniform grid in the BZ, and the integral is evaluated as a finite sum:
\begin{multline}
\label{eq_GW_expl_corr}
   \Sigma^c_{n\kv}(\omega) =  -\frac{1}{\Omega}  \sum\limits_m \sum\limits_{\mathbf{G}\mathbf{G}'}
    \int\frac{d\omega'}{2 \pi i} e^{i \omega' \eta} \\ \sum_{\qv} g_{\mathbf{G}\mathbf{G}'}^{n \kv}(\qv,\omega,\omega') \, \overline{W}^{c}_{\mathbf{G}\mathbf{G}'}(\qv,\omega').
\end{multline}
 In the sum, we define $\overline{W}^{c}_{\mathbf{G}\mathbf{G}'}$ as the average of 
$W^{c}_{\mathbf{G}\mathbf{G'}}$ within the mini-BZ of size $D_{\Gamma}$ centered around each $\qv$ points of the Monkhorst-Pack grid:
\begin{equation}
\label{Eq_W_av}
\overline{W}^{c}_{\mathbf{G}\mathbf{G'}}(\qv,\omega) = \frac{1}{D_{\Gamma}}\int\limits_{D_{\Gamma}}\! \! \! d\qv' \ W^{c}_{\mathbf{G}\mathbf{G'}}(\qv+\qv',\omega).
\end{equation}
The inclusion of this average is particularly important for systems in which the dielectric matrix and the bare Coulomb potential vary on a similar scale for $\qv \to 0$ and therefore $W^c$ cannot be approximated as piecewise constant. 

The integrals of \cref{Eq_W_av} can be conveniently evaluated via a Monte Carlo method. However, $W^c$ is calculated only at the the center of each cell, and its analytic $\qv$-dependence is unknown. In the W-av method, $W^c$ is extended beyond the grid points to the full BZ by interpolating an auxiliary function, $f_{\mathbf{G}\mathbf{G'}}(\qv,\omega)$, whose variation with $\qv$ is smoother than $W^c$:

\begin{align}
\label{W_rec}
\! \! W^{c}_{\mathbf{G}\mathbf{G'}}(\qv,\omega) &= \frac{V_{\mathbf{G}}(\qv)f_{\mathbf{G}\mathbf{G'}}(\qv,\omega)V_{\mathbf{G'}}(\qv)}{1- \sqrt{ V_{\mathbf{G}}(\qv)} f_{\mathbf{G}\mathbf{G'}}(\qv,\omega) \sqrt{ V_{\mathbf{G'}}(\qv)} },
\\[6pt]
\label{eq.f}
\! \! f_{\mathbf{G}\mathbf{G'}}(\qv,\omega) &=\frac{\chi_{\mathbf{G}\mathbf{G'}}(\qv,\omega) }{1+ \sqrt{ V_{\mathbf{G}}(\qv)} \chi_{\mathbf{G}\mathbf{G'}}(\qv,\omega) \sqrt{ V_{\mathbf{G'}}(\qv)}}.
\end{align}
\begin{figure}[h]
    \centering
    \includegraphics[width=0.9\linewidth]{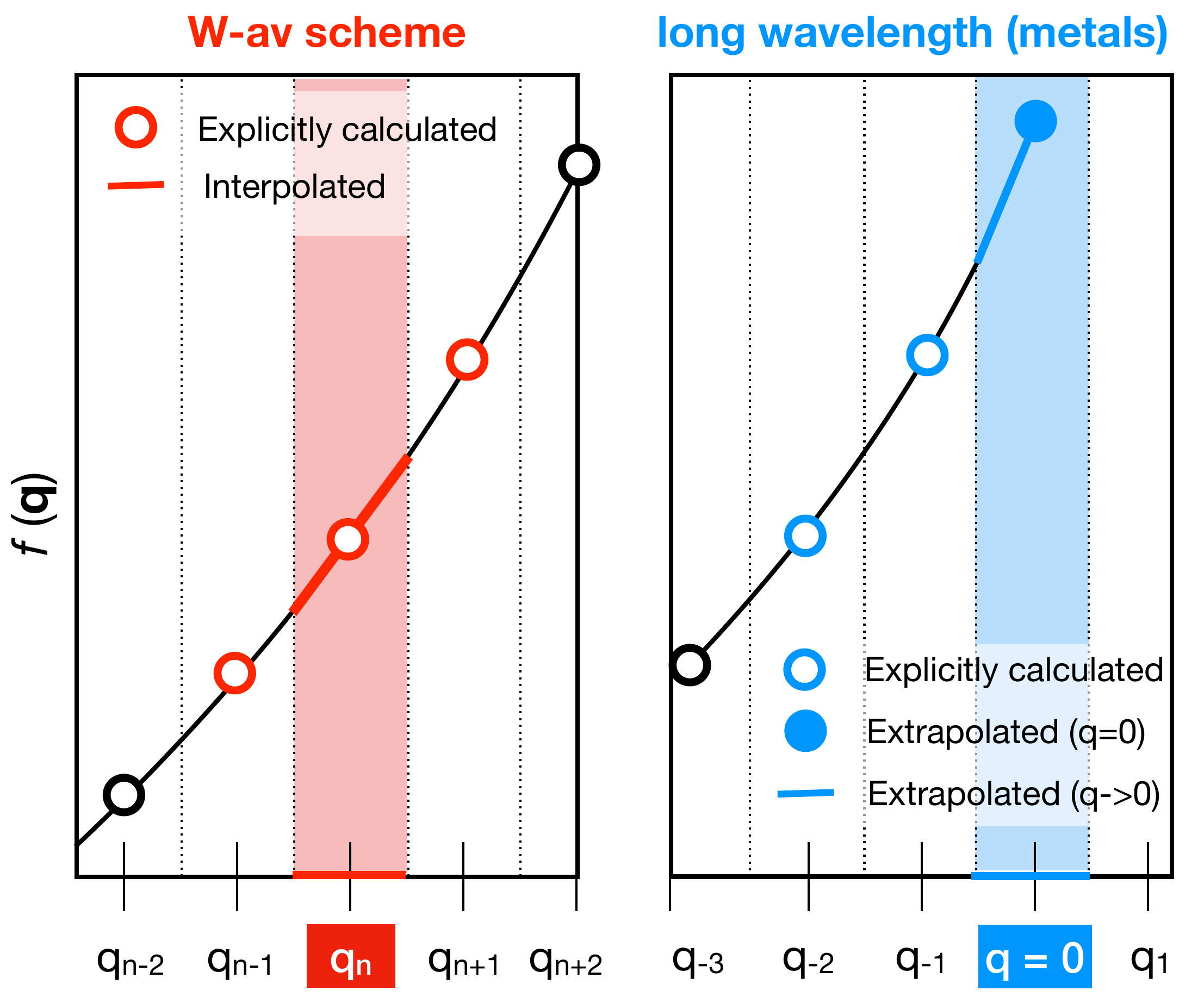}
    \caption{Scheme illustrating the interpolation ($q\neq0$) and the long-wavelength extrapolation ($q=0$) procedures used within the W-av method to extend the $f$ function to all points in the BZ zone, beyond the k-grid points.
    }
    \label{fig:scheme}
\end{figure}
As illustrated in the left panel of Fig.~\ref{fig:scheme}, within each cell, $f_{\mathbf{G}\mathbf{G'}}(\qv,\omega)$ is interpolated in the whole cell using its values at the center of that same cell and of the neighbouring cells, through a quadratic polynomial in  $\qv'$ (here, $\qv$ and $\qv'$ are written in reciprocal lattice coordinates):
\begin{multline}
\label{Eq_f_fit}
f_{\mathbf{G}\mathbf{G'}}(\qv+\qv) \equiv f_{\mathbf{G}\mathbf{G'}}(\qv) +  \bar{f}_{\mathbf{G}\mathbf{G'}}(\qv) \cdot \qv'\\
+\qv'\cdot \bar{\bar{f}}_{\mathbf{G}\mathbf{G'}}(\qv)\cdot \qv' ,
\end{multline}
\begin{equation}\label{Eq_f_coeff_v}
\bar{f}_{\mathbf{G}\mathbf{G'}}(\qv) = \begin{bmatrix}
f^{1}_{\mathbf{G}\mathbf{G'}}(\qv) &
f^{2}_{\mathbf{G}\mathbf{G'}}(\qv)
\end{bmatrix},
\end{equation}
\begin{equation}\label{Eq_f_coeff_m}
\bar{\bar{f}}_{\mathbf{G}\mathbf{G'}}(\qv) = \begin{bmatrix}
f^{11}_{\mathbf{G}\mathbf{G'}}(\qv) & 0\\
0 & f^{22}_{\mathbf{G}\mathbf{G'}}(\qv) 
\end{bmatrix}.
\end{equation}
For simplicity, the interpolation is performed using a bilinear form for $\bar{\bar{f}}_{\mathbf{G}\mathbf{G'}}(\qv)$.  
The analytic expression of $W^c$, determined for each cell, is then used to calculate, first the averaged potential in \cref{Eq_W_av}, and finally the correlation part of the self-energy in \cref{eq_GW_expl_corr}. For 2D semiconductors, the W-av method provides a remarkable convergence acceleration with respect to the $\kv$-grid size~\cite{Guandalini2023npjCM, GuandaliniLeon_2024}.

\subsection{W-av method for metals}
\label{methods:Wav_metals}

The W-av method developed for semiconductors is not directly transferable to systems with metallic screening, since the contribution of intra-band transitions in the polarizability makes it difficult to compute $W^c$ at the center of the $\qv=0$ cell. 
The RPA polarizability is here calculated according to Adler and Wiser~\cite{Adler1962PhysRev,Wiser1963PhysRev}:
\begin{multline}
\label{eq.polarizability} 
\chi^0_{\mathbf{G}\mathbf{G'}}(\qv,\omega) = 2 \sum_{n,m} \int\frac{d\kv}{(2\pi)^3}   
\, \left[\mathfrak{f}(\varepsilon_{n \kv}) - \mathfrak{f}(\varepsilon_{m \kv+\qv})\right] \times \quad
\\ 
\frac{ \rho_{nm}(\kv,\qv,\mathbf{G}) \, \rho^{*}_{nm}(\kv,\qv,\mathbf{G'}) }{\omega + \varepsilon_{n \kv} - \varepsilon_{m \kv+\qv} + i \eta \ \textrm{sign}(\mathfrak{f}(\varepsilon_{n \kv}) - \mathfrak{f}(\varepsilon_{m \kv + \qv })) },
\end{multline}
where $\mathfrak{f}(\varepsilon_{n \kv})$ is the Fermi-Dirac distribution, the indices $n$ and $m$ run on all valence and conduction bands. The main difficulty in the calculation of the polarizability lies in the limit $| \qv|\rightarrow 0$, corresponding to electronic transitions of vanishing energy. Due to the $\qv$ discretization, the intra-band transitions must be treated either as a transition between the same state, with the same $\kv$ and energy, or between different states, immediately below and above the Fermi surface, resulting in a spurious energy gap. It is standard practice to simply neglect these long-wavelength intraband contributions, decreasing their weight by increasing the number of $\kv$ points. 
This means that, for metals, $W^c$, and thus $f$, is not computed for $\qv=0$, i.e., at the center cell $D_{\Gamma}$, hindering the interpolation described in the previous Section. 
\begingroup

\setlength{\tabcolsep}{4pt} 

\begin{table}[ht!]
    \centering
    \renewcommand{\arraystretch}{2}%
    \begin{tabular}{ccc}
    \hline\hline 
         3D           & static  & dynamic \\[-12pt]
         metal  & ($\omega \ll q^2$) & ($\omega \gg q^2$) \\[6pt]
         \hline
         \hline
         &&\\[-18pt]
          $f_{00}$ & $- \frac{ \kappa^{M \ 2}_{TF}}{4 \pi }+ o(\qv) $   & $\frac{q^2}{4 \pi} \frac{\omega_p^2(\hat{\qv})}{\omega(\omega+ i \gamma)}+o(\qv^3)$\\[4pt]
         \hline
         &&\\[-18pt]
         W$^c_{00}$ & $-\frac{4 \pi}{q^2}\frac{\kappa^{M \ 2}_{TF}}{q^2+\kappa^{M \ 2}_{TF}}  + o(\qv^{-1})$   & $\frac{4 \pi}{q^2}\frac{\omega_p^2(\hat{\qv})}{\omega(\omega+ i \gamma)-\omega_p^2(\hat{\qv})} + o(\qv^{-1})$\\[6pt]
         \hline         
         \hline
          $f_{\mathbf{G}0}$ & $q^2+o(\boldsymbol{q}^3)$   & $\boldsymbol{q}+o(\qv^2)$\\[6pt]
                   \hline
         W$^c_{\mathbf{G}0}$ & $ q^0 +o(\boldsymbol{q}) $  & $\qv^{-1}+o(\qv^0)$\\[6pt]
         \hline  \hline       
    \end{tabular}
 \caption{Summary of the long-wavelength $\boldsymbol{q}$-dependence of the head and wing elements in a 3D metal for the auxiliary functions $f$ (see~\cref{eq.f}) and the correlation part of the screened Coulomb potential $W^c=W-V$ (see~\cref{eq.W}). In the table, $\kappa^{M}_{TF}$ indicates the macroscopic Thomas-Fermi wavevector,
 while $\omega_p(\hat{\boldsymbol{q}})$ indicates the plasmon frequency that may be directional. Details on $\kappa^{M}_{TF}$ and $\omega_p(\hat{\boldsymbol{q}})$ are provided in Eq.~(S39) and Eq.~(S42) of the SI.
  \label{table:q3D} 
 }
\end{table}

\endgroup

\begingroup

\setlength{\tabcolsep}{6pt} 

\begin{table}[hb!]
    \centering
    \renewcommand{\arraystretch}{2}%
    \begin{tabular}{ccc}
\hline\hline
        2D    & static  & dynamic \\[-12pt]
        metal  & ($\omega \ll q^2$) & ($\omega \gg q^2$) \\[6pt]
         \hline
         \hline
         &&\\[-18pt]
          $f_{00}$ & $-\frac{ \kappa^{M}_{TF}}{2 \pi L} + o(\qv)$  & $\frac{q}{2 \pi L} \frac{\omega_p^2(\boldsymbol{q})}{\omega(\omega+ i \gamma)}+o(\qv^3)$\\[4pt]
         \hline&&\\[-18pt]
         W$^c_{00}$ & $-\frac{2 \pi L}{q}\frac{\kappa^{M}_{TF}}{q+\kappa^{M}_{TF}} + o(\qv^0)$ & $\frac{2 \pi L}{q}\frac{\omega_p^2(\qv)}{\omega(\omega+ i \gamma)-\omega_p^2(\qv)} + o(\qv)$\\[6pt]
         \hline         
         \hline
$f_{\mathbf{G}_{\parallel}0}$ & $q+o(\qv^2)$ & $\qv +o(\qv^2)$ \\ 
$f_{\mathbf{G}_{z}0}$ & $q+o(\qv^2)$ &     $\qv^2 +o(\qv^3)$ \\[6pt]
          \hline
$W^c_{\mathbf{G}_{\parallel}0}$ & $q^0+o(\qv)$ & $\hat{\qv}+o(\qv)$ \\
$W^c_{\mathbf{G}_{z}^{\text{odd}}0}$ & $q^0+o(\qv)$& $\qv+o(\qv^2)$ \\ 
$W^c_{\mathbf{G}_{z}^{\text{even}}0}$ & $q+o(\qv^{2})$ & $\qv^2+o(\qv^{3})$ \\[6pt]
\hline \hline     
    \end{tabular}
 \caption{Summary of the long-wavelength $\boldsymbol{q}$-dependence of the head and wing elements in a 3D metal for the auxiliary functions $f$ (see~\cref{eq.f}) and the correlation part of the screened Coulomb potential $W^c=W-V$ (see~\cref{eq.W}). Here, $\kappa^{M}_{TF}$ is the macroscopic Thomas-Fermi wavevector,
 and $\omega_p(\boldsymbol{q})$ is the intraband plasmon of a 2D metal. Details on $\kappa^{M}_{TF}$ and $\omega_p(\boldsymbol{q})$ are provided in Eq.~(S63) and Eq.~(S65) of the SI.
 The wing elements are split in three different types: $\mathbf{G}_{\parallel}$ indicates the RL vectors with in-plane contribution, $\mathbf{G}_z^{\text{odd}}=n\pi/L$ out-of-plane RL vectors with $n$ odd and $\mathbf{G}_z^{\text{even}}=n\pi/L$ out-of-plane RL vectors with $n$ even.
 \label{table:q2D}}
\end{table}

\endgroup
Within the W-av framework, this is addressed by replacing, in the long-wavelength limit, the interpolation  by a physics-informed extrapolation of the behavior at the limit of the auxiliary functions $f$ of~\cref{eq.f}~\cite{note_yambo_vars}, as illustrated in the right panel of Fig.~\ref{fig:scheme}. 
The function $f$ within the whole $D_{\Gamma}$ cell is now determined from an extrapolation using the $f$ values computed at the center of the first- and second-order neighbours of the $\qv = 0$ element along the reciprocal-cell vectors. 
While the extrapolation is formally similar to \cref{Eq_f_fit}, the choice of the polynomial type depends on the specific ${\mathbf{G},\mathbf{G'}}$ components. 
For the body elements ($\mathbf{G}\neq0$, $\mathbf{G'}\neq 0$), the extrapolation is done  with a standard quadratic polynomial, since the bare Coulomb potential behaves regularly. 
In contrast, for the head ($\mathbf{G}=\mathbf{G'}=0$) and wings ($\mathbf{G}=0$, $\mathbf{G'}\neq 0$) terms, we rely on the relations between $f_{\mathbf{G}\mathbf{G'}}(\qv,\omega)$ and $\chi^0_{\mathbf{G}\mathbf{G'}}(\qv,\omega)$ to provide the correct small-$\qv$ functional form. We refer to Section S I of the Supplementary Information (SI) for the derivation of the relations, which lead to the following expressions:
\begin{eqnarray}
f_{00}(\qv,\omega) &=& \chi^0_{00}(\qv,\omega) -\sum_{\mathbf{G}\mathbf{G}'} \chi^0_{0\mathbf{G}}(\qv,\omega)
\nonumber
\\ 
&\times&  \sqrt{V_{\mathbf{G}}(\qv)}  B^{-1}_{\mathbf{G}\mathbf{G}'} \sqrt{V_{\mathbf{G'}}(\qv)} \, \chi^0_{\mathbf{G'}0}(\qv,\omega),
\\[7pt] \nonumber
f_{\mathbf{G}0}(\qv,\omega) &=& - B^{-1}_{\mathbf{G}\mathbf{G'}} \, \chi^{0}_{\mathbf{G'}0}(\qv,\omega) / \Big[ 1- V_0(\qv) f_{00}(\qv,\omega) 
\\ 
&+&  \sqrt{ V_{\mathbf{G}}(\qv) } \, B^{-1}_{\mathbf{G}\mathbf{G'}}\chi^{0}_{\mathbf{G'}0}(\qv,\omega ) \sqrt{ V_{0}(\qv)} \Big],
\end{eqnarray}
where $B^{-1}_{\mathbf{G}\mathbf{G'}}$ is the body of the symmetrized dielectric matrix, thus defined only for $\mathbf{G}\neq 0$ and $\mathbf{G'}\neq 0$, given by
\begin{equation}
B_{\mathbf{G}\mathbf{G'}}=\delta_{\mathbf{G}\mathbf{G'}} - \sqrt{V_{\mathbf{G}}(\qv)} \, \chi^0_{\mathbf{G}\mathbf{G'}}(\qv,\omega) \sqrt{V_{\mathbf{G'}}(\qv)} .
\end{equation}
In addition to the $\mathbf{G}, \mathbf{G}'$ components, a distinction must be made between the static and dynamical regime. Similarly to $\chi^0_{\mathbf{G}\mathbf{G'}}(\qv,\omega)$, the long-range behavior of $f_{\mathbf{G}\mathbf{G'}}(\qv,\omega)$ is different in the static ($\omega=0$) and dynamic ($\omega\neq0$) cases. This makes the W-av scheme fully compatible with a full frequency treatment, here accomplished within the MPA. Nevertheless, the W-av approach is also compatible with PPA schemes. Moreover, $V_{\mathbf{G}}(\qv)$ is truncated in the case of 2D systems~\cite{ismail2006truncation,Rozzi2006PRB}.

Comparing with the existing literature, we emphasize that, although there are other schemes to treat the $W(\qv\rightarrow 0)$, those treatments generally do not include dynamical effects~\cite{Deslippe2012CPC}, or include them only on head term of the $W$, though recurring to an increase of the density grid~\cite{da2020universal}. 
Another difference is in the $q \to 0$ extrapolating procedure, which typically resorts just to the leading order, and it is based on a single small $q$-value, while here we consider all the neighbours of the $q=0$ point of the Monkhorst-Pack grid. This allows the W-av method to be more general and applicable also to anisotropic materials both in the static and dynamic limit.
A detailed derivation and discussion the all the functional forms of $f_{\mathbf{G}\mathbf{G'}}(\qv,\omega)$ can be found in Section S I of the SI. The head and wings components of $f$ and of $W^c$ are summarized in~\cref{table:q3D,table:q2D}. Note that, once summed to the Coulomb potential $V$, the expansion of $W^c$ recovers the known long-wavelength expression of $W$. For a better understanding of the expressions in the tables, some key aspects of the derivation are briefly commented in the following paragraph.

The static term of $f$ is related to the Thomas-Fermi wave vector of the material which is known~\cite{Giuliani-Vignale2005book} to be constant and isotropic at the lowest order, being related to the density of states on the Fermi surface. For this reasons, $f_{00}(\qv \to 0,0)$ is taken as the average over all directions of the $\kv$ grid.
In contrast, the finite frequency terms are in general anisotropic, and therefore the $\qv$ direction must be taken into account. When the $f_{\mathbf{G}\mathbf{G'}}(\qv \to 0, \omega)$ expansion to the lowest order is quadratic, the parameterization in Eq.~\eqref{Eq_f_fit} is expanded to include third-order terms. 
For the 3D case at finite frequency, we also relax the bilinear approximation in the quadratic terms of $f_{00}(\qv, \omega)$, both for the $\qv \to 0$ extrapolation and for the neighbouring cells, and thus the non-diagonal terms in Eq.~\eqref{Eq_f_coeff_m} are allowed to take nonzero values. By including these non-diagonal terms the $q^2$ long-range behavior that matches the Coulomb potential $\sim q^{-2}$ dispersion is strictly enforced, thus avoiding first-order terms that introduce numerical noise. The non-diagonal terms are determined by including in the extrapolation neighbouring $\qv$ points reached by combinations of two non-collinear reciprocal lattice vectors.

\subsection{Band gap renormalization in doped 2D semiconductors}
\label{section:bandgap_renorm_2D}
%
\begin{figure*}[t!]
    \centering
    \includegraphics[width=\textwidth, trim={0.2cm 0.5cm 0.8cm 0.2cm },clip]{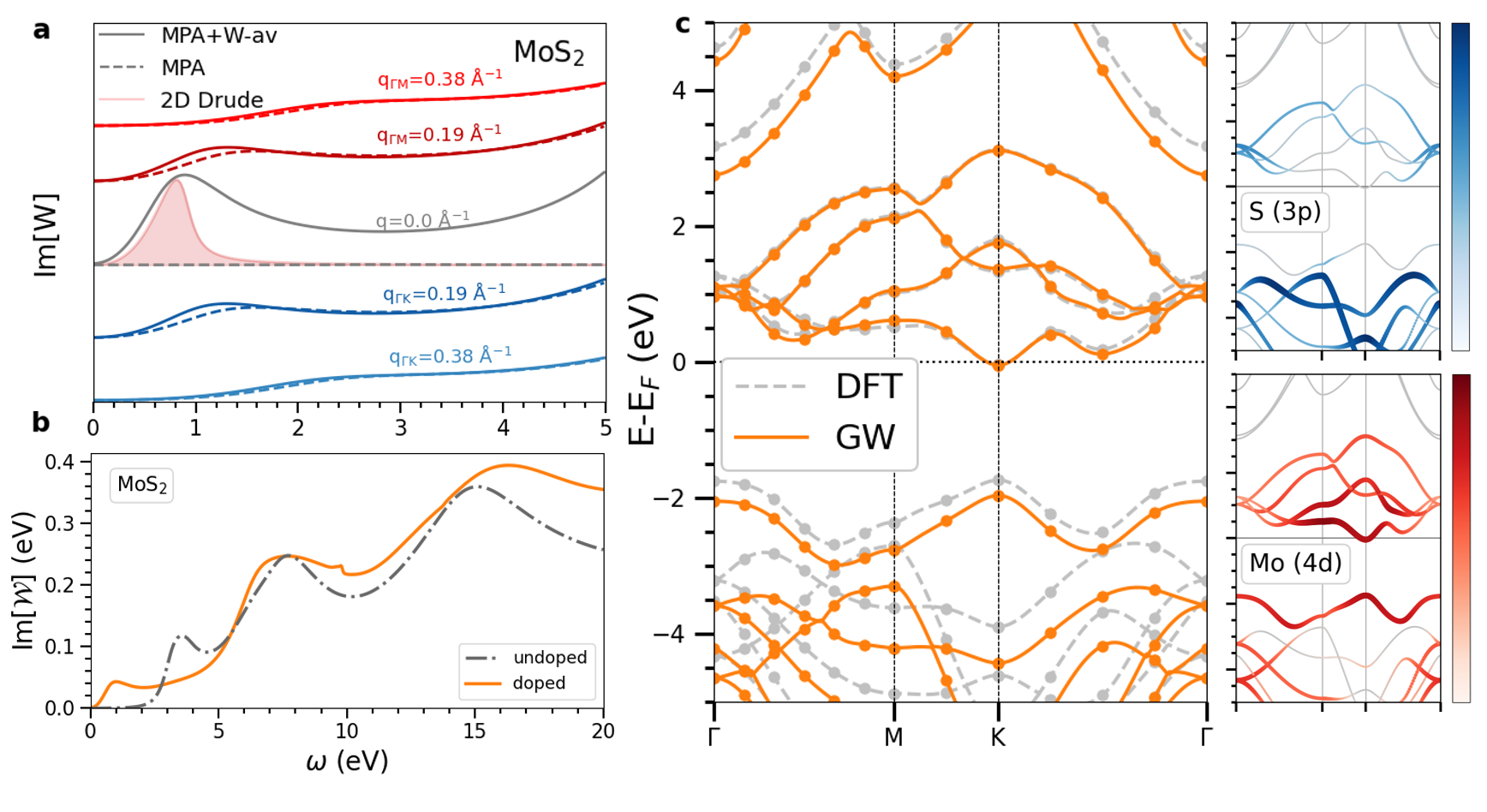}  
    \caption{$n$ doped MoS$_2$ with $\rho$=4.56$\times 10^{13}$ el/cm$^{2}$ (0.04 el/cell): \textbf{a)} Imaginary part of $W_{00}(\qv,\omega)$ with respect to the frequency $\omega$ in the low-energy region at small momenta $q$. For reference, $\Gamma$M=$1.14$~\AA$^{-1}$ and $\Gamma$K=$1.32$~\AA$^{-1}$. The curves are on the same scale and have been shifted for visualization purposes.
    \textbf{b)} Imaginary part of the screened Coulomb potential $\mathcal{W}(\omega)$ [see \cref{eq.Wreal}] compared to the undoped case. The calculations are run using the MPA+W-av method. Results are shown with an additional broadening of 0.01 eV for visualization purposes.      
    \textbf{c)} DFT (grey) and $GW$ (orange) band structure along with the DFT projected density of states of the relevant orbital composition in the first valence and conduction bands. The dots indicate the numerical points of the $GW$ calculation on a 12x12 $\kv$-grid. The band structure within the points has been determined as discussed in the computational details. The dotted-horizontal line signals the Fermi energy.}
    \label{fig:MoS2_united}
\end{figure*}
As anticipated, the reduced screening of 2D materials causes the energy levels to be easily affected by external factors governing the dielectric screening. Doping, for instance, can cause a significant renormalization of the quasi-particle gaps of semiconductors~\cite{Ugeda2014,Mak2012,Ross2013,Huser2013,Louie2021NatMater}. However, the first principles description of these screening effects is challenging, particularly for low doping densities, for which the Fermi surfaces can be very small. On the other hand, the small doping regime is also the most relevant for realistic experiments and devices.
In this Section, we examine the effect of doping in the gap and band structure of monolayer MoS$_2$ and graphene. The present results were computed within the $GW$ approximation, obtained using the W-av method generalized to metals, described above. The discussion of the convergence speed-up enabled by W-av can be found in Section~\nameref{sec:3}. \\

We start by analyzing the case of doped MoS$_2$. It is instructive to
compare the screened potential, computed with and without the W-av method, focusing on the low-frequency region, where the free carriers plasmon is expected. 
Figure~\ref{fig:MoS2_united}\textbf{a} shows the first $\qv$-components of $\overline{W}^{c}_{00}$ for doped monolayer MoS$_2$ with a $n$-type doping of $\rho=4.56\times 10^{13}$ el/cm$^{2}$ (0.04 electrons per cell), evaluated along the two main crystallographic directions on a $12\times12$ grid. For finite-$\qv$ elements, the W-av and standard calculations yield only minor differences, with the intraband plasmon correctly reproduced in both cases. In contrast, in the long-wavelength limit, the plasmon peak is correctly described only by the W-av method, as shown by the solid grey curve.  Notice that, even if the intra-band contribution to W in 2D  disperses as $\omega_p \sim \sqrt{q}$~\cite{stern1967,hwang2007dielectric}, the peak resulting from a MP discretization lies at finite energy also for $q=0$, as W is averaged within each cell [see~\cref{Eq_W_av}].
To better understand this behavior, we considered a simple 2D Drude model, with a plasmon frequency given by the effective mass computed from the parabolic band at K ($m^*=0.424 m_e$), using the expression $\omega_p=\sqrt{\frac{2 \pi \rho }{m^*}q}$~\cite{stern1967}.
The frequency of the integrated model W, plotted as an orange shaded peak of custom width in Fig.~\ref{fig:MoS2_united}\textbf{a}, agrees well with the W-av result and confirms the intraband origin of the low energy peak.

Figure~\ref{fig:MoS2_united}\textbf{b} shows the momentum-integrated screened interaction for the doped and pristine systems, defined as a discrete sum~\footnote{The summation is here restricted to the in-plane momenta $\qv$ of the Monkhorst–Pack grid. As a consequence, the absolute value of $\mathcal{W}(\omega)$ depends on the supercell used, which is the same for all doping values. The dependency on the supercell is removed once taken the sum over the $G_z$ vectors.}  over all the $\qv$ points of the $\overline{W}^{c}_{00}$ elements,
\begin{equation}
\label{eq.Wreal}
\mathcal{W}(\omega)= \frac{1}{\Omega}
\sum_{\qv}
\overline{W}^c_{00}(\qv,\omega).
\end{equation}
We focus in particular on the imaginary part of $\mathcal{W}(\omega)$, that provides indication of the collective charged electronic excitations.
In the doped case (orange curve), $\mathcal{W}(\omega)$ presents features spread across a large energy range. The higher-energy structures, associated with interband transitions, occur at energies comparable to those of the pristine system (gray curve), although the different electronic occupations renormalize their intensities. For example, around 3.5~eV, the distinct peak present in the pristine case is almost suppressed for the  doped system.
The low-energy intra-band plasmon, discussed in the previous paragraph, appears as a distinct peak, absent in the pristine case. Despite being located in different energy regions, the W-av method accurately describes both contributions (see also Fig. \ref{fig:comparison_W_2D}\textbf{a} for reference).

In~\cref{fig:MoS2_united}\textbf{c} we present the band structure of MoS$_2$ for the same level of $n$ doping, as obtained from DFT (in silver) and $GW$ calculations (in orange) performed using the W-av method for metals. The electron doping results in a small pocket of free carriers around the $K$ point. Compared to the undoped case~\cite{Guandalini2023npjCM}, the calculated $GW$ corrections are still significant but noticeably reduced. For instance, the $GW$ correction to the valence-conduction gap at $K$ reduces from a value of 0.8~eV in the undoped case to 0.2~eV for the doping considered. Once accounted for the realignment of the Fermi level, we find that the $GW$ corrections are particularly small for the metallic band and for the first conduction bands above the Fermi level, ranging around few tens of meV with a notable exception  around $\Gamma$ where they are larger than 100 meV, and they progressively increase for states further away from Fermi. At the same time, the energy separation between the metallic and valence bands increases throughout the Brillouin zone, without significantly changing the bands dispersion. 
\begin{figure}[b!]
    \includegraphics[width=0.48\textwidth, trim={0.37cm 0.42cm 0.0cm 0.0cm },clip]{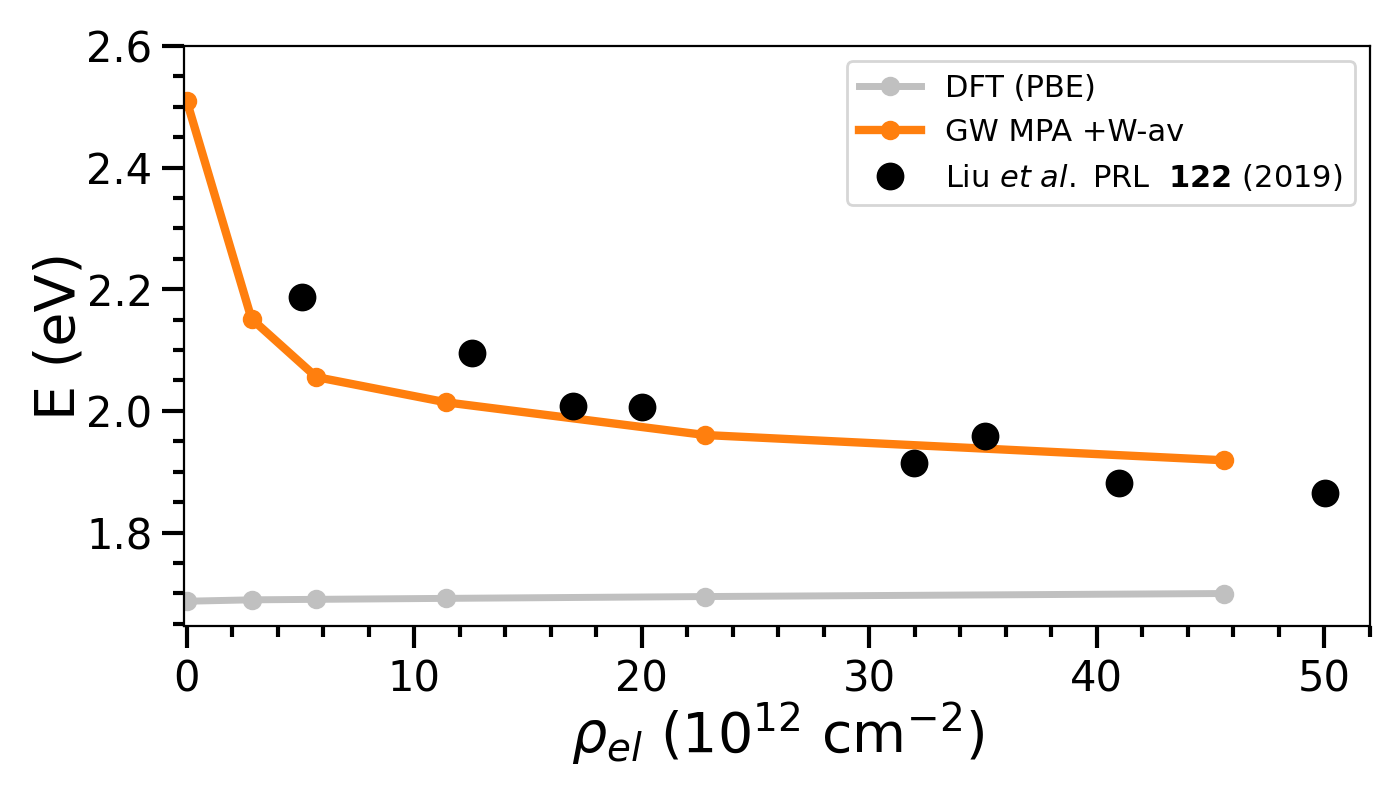}
    \caption{MoS$_{2}$: Band gap at the $K$ point at different levels of $n$ doping, as obtained from DFT and $GW$ calculations in this work compared to TR-ARPES experimental results~\cite{LiuMoS2}.}
    \label{fig:QPcorr_MoS2}
\end{figure}
\begin{figure*}[t!] 
    \includegraphics[width=\textwidth, trim={0.13cm 0.5cm 0.13cm 0.0cm },clip]{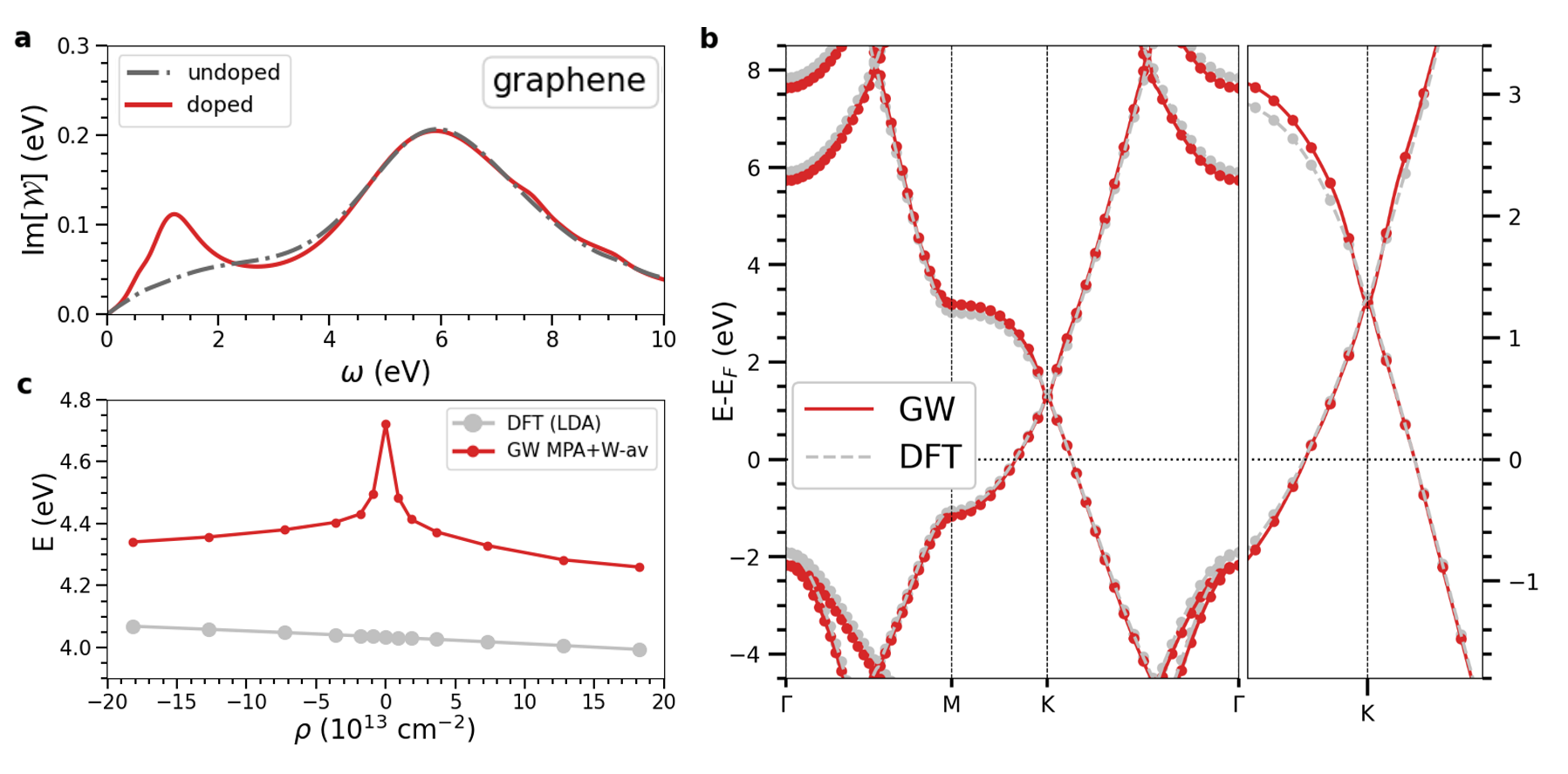}     
    \caption{Graphene $p$ doped with $\rho$=-19.09$\times 10^{13}$~el/cm$^{2}$ (-0.1 el/cell): \textbf{a)} Imaginary part of the screened Coulomb potential $\mathcal{W}(\omega)$.  Results are shown with an additional broadening of 0.2 eV for visualization purposes. The rather large value of broadening is discussed more in detail in Section S IV of the SI. \textbf{b)}  DFT (grey) and $GW$ (red) band structure. The dots indicate the numerical points on a 60x60 $\kv$-grid. The band structure within the points has been determined as discussed in the computational details. \textbf{c)} Band gap at the M point at different levels of graphene doping as found from DFT and $GW$ calculations using the W-av method. \label{fig:graph_united}} 
\end{figure*}
The resulting asymmetry of the $GW$ corrections below and above the Fermi level can be explained in terms of the orbital character of the bands. We show the projected bands structure at DFT level of the two most relevant orbital contributions: the S-$3p$ and Mo-$4d$ orbitals. The first conduction bands show a predominantly $d$ character, but in the vicinity of the $\Gamma$ point there is a significant $p$-orbital contribution. Consequently, once particle-number conservation is accounted for in~\cref{eq.gw_ex}, the $GW$ corrections turn out small and nearly uniform for all the $d$-character conduction bands. The notable exception is around $\Gamma$, where the $p$-character leads to larger corrections amounting to approximately 200 meV. A similar argument applies also to the valence band, which exhibits a mixed $d$–$p$ character across the Brillouin zone. As shown in a previous work by our group~\cite{Leon_2023}, $d$ states are particularly sensitive to the treatment of the frequency dependence and require a proper handling of intra-band contributions. These contributions are included in W-av@MPA making it able to properly describe the band-gap widening at the $K$ point. 
In~\cref{fig:QPcorr_MoS2} we show the dependence of the band gap at the K point on the electronic doping, computed at the DFT and $GW$ levels. While the DFT results show an almost constant band gap, the $GW$ band gap decreases sharply when departing from the charge neutrality point, in agreement with previous results~\cite{AsgariMoS2}. As doping increases, the band gap progressively flattens. Our findings are in very good agreement with TR-ARPES measurements~\cite{LiuMoS2} carried out on MoS$_2$ suspended on a dielectric, which provides close-to-intrinsic renormalization of the band-gap.
The fully ab initio treatment of both the dynamical response, through MPA, and the long wavelength limit, through W-av, results in a improved agreement with experiment with respect to previously reported $GW$ results \cite{Liang2015, Gao2017}. 
Previously, the description of low-energy intraband contributions in 2D system was achieved through the inclusion, only in the head matrix element, of an additional model intraband pole into a Hybersten and Louie PPA response~\cite{Liang2015,Gao2017}.
As the W-av@MPA being used here samples a broad but finite region in the frequency space, it does not comply by design with the $f$-sum rule as, e.g. the Hybersten and Louie PPA model. However, MPA allows for a more consistent description of the response over a broad energy range when compared with single-pole or two-poles models, comprised of inter- and intra-band features, which are described in the entire dielectric matrix fully \textit{ab initio}. \\       

We now examine the case of doped graphene, first considering a specific $p$ doping of $\rho$=-19.09$\times 10^{13}$~el/cm$^{2}$ (-0.1 electrons per cell). Figure~\ref{fig:graph_united}\textbf{a} shows the momentum-integrated screened interaction $\mathcal{W}(\omega)$ for pristine graphene in grey and the $p$ doped one in red. At low energies, the presence of free carriers has a noticeable effect: the $e$-$h$ onset characteristic of undoped graphene is hardly visible, due to the presence of the intraband (Dirac) plasmon.  
At higher energies, on the other hand, $\mathcal{W}(\omega)$ is mostly unchanged. In fact, this energy range hosts the $\pi$ plasmon that originates from interband transitions close to the $M$ point, that are not affected by doping. Previous calculations of doped graphene~\cite{albertoTB} show that the RPA response function obtained on top of KS-DFT electronic structure provides a reasonable dispersion of the $\pi$-plasmon as compared to $GW$+BSE simulations, contrarily to the undoped case~\cite{guandalini_2023}.

Figure~\ref{fig:graph_united}\textbf{b} shows the band structure of $p$ doped graphene with -0.1 electrons per cell, as obtained from DFT and $GW$ calculations using the W-av method for metals. The $p$ doping shifts the Fermi level approximately $1.34$ eV below the Dirac point, close to the bottom of the linear dispersion region. While the $GW$ corrections are generally smaller than those observed in doped MoS$_2$, they follow a similar trend, becoming larger for states further away from the Fermi level and in correspondence to the  bands flattening. The $GW$ corrections also change the slope of the Dirac cones, asymmetrically with respect to the Dirac point, due to doping. In fact, the $GW$ corrections at the Fermi energy are bound to be zero, but they are also very small at the Dirac point, due to the zero density of states. As a consequence, in the upper cone, close to the Dirac point, the steepness increases, on average, from $0.83\times10^6$~m/s (DFT) to $0.99\times10^6$~m/s ($GW$), whereas, in the lower part, it slightly decreases from $0.84\times10^6$~m/s (DFT) to $0.81\times10^6$~m/s ($GW$). Outside the linear dispersion region along the $\Gamma$M direction, both in the upper and lower cone, the $GW$ corrections  become even more appreciable, as shown in the zoomed band structure of Fig.~\ref{fig:graph_united}\textbf{b}. In the upper cone, there is also the formation of a kink along K$\Gamma$, at around 2.5~eV, consistently with previous studies~\cite{Jakovac2024}.

Finally, we evaluate the quasi-particle energy difference between the higher occupied and lowest unoccupied state at the $M$ point for different doping levels, shown in ~\cref{fig:graph_united}\textbf{c}. The general pattern agrees with earlier literature~\cite{attaccalite2010doped}, indicating a sharp decrease in QP energies upon charge injection into the system, but the magnitude and the profile of the renormalization are significantly changed. 
This difference is due to the improved description of W-av@MPA in $\kv$ and frequency space to describe the metallicity of the system, while previously, no specific treatment of the  intra-band contributions was adopted, and the frequency dependence was described with a simple PPA.

\begin{figure*}[t!]
    \centering  
    \includegraphics[width=\textwidth,trim={0cm 0.0cm 0.0cm 0.0cm },clip]{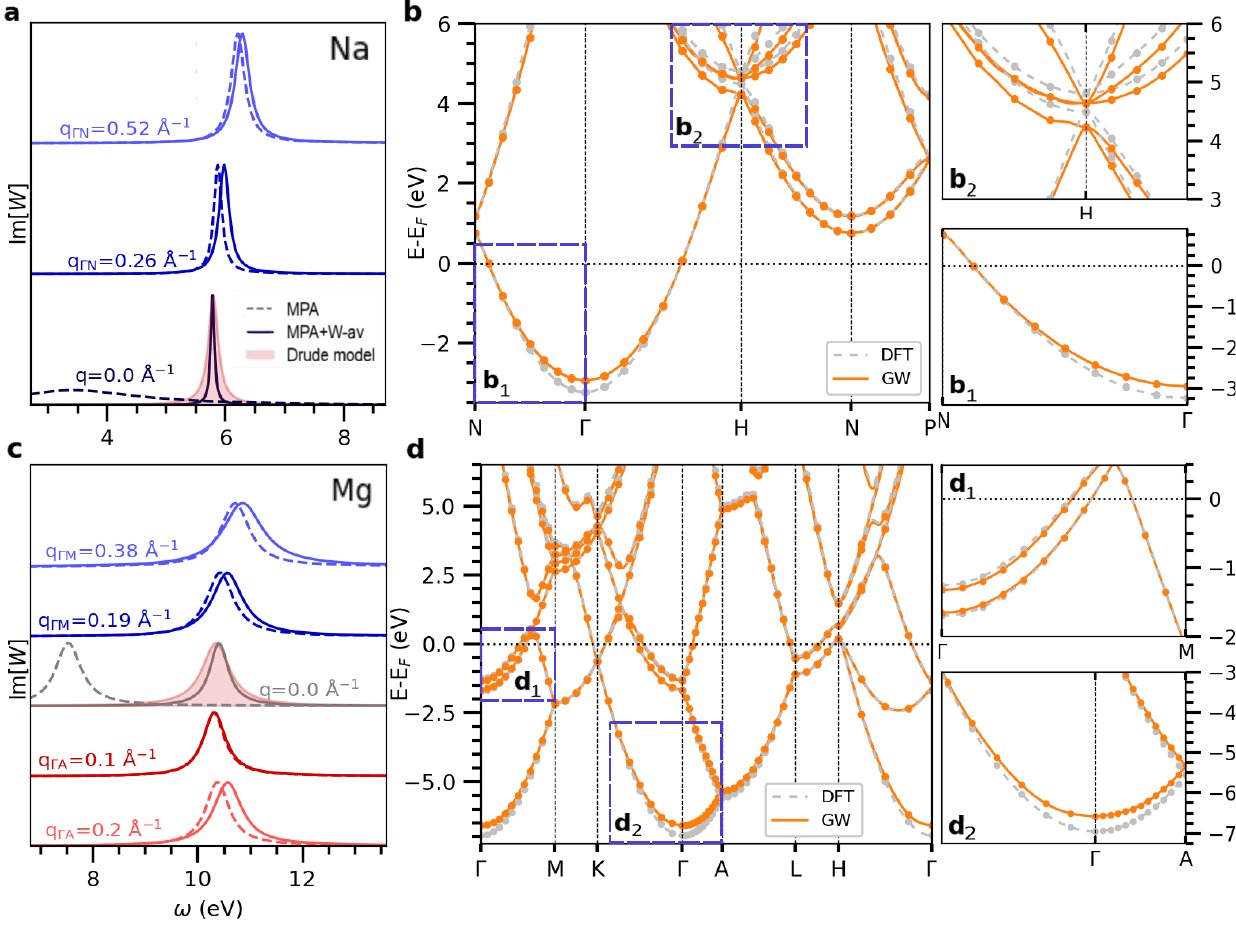}
    \caption{Plots for \textbf{a)}-\textbf{b)} bulk Na and \textbf{c)}-\textbf{d)} bulk Mg showing, respectively, the imaginary part of $W_{00}(\qv,\omega)$ with respect to frequency $\omega$ at small momenta $\qv$ (left panels) and the DFT (grey) and $GW$ (orange) band structures (right panels). The imaginary part of $W_{00}(\qv,\omega)$ (\textbf{a}-\textbf{c}) is shown in the region of the bulk plasmon, represented by a Lorentzian function with a damping of 0.1~eV and centered at 5.8~eV for Na and 10.35~eV for Mg. Results obtained with (solid lines) and without (dashed) the W-av method are compared. In case of Mg, both in-plane and out-of-plane wavevectors of the reciprocal HCP lattice are included. \textbf{b}-\textbf{d)} The band structures are shown together with the calculated points on a $16\times16\times16$ $\kv$-grid for Na and a $24\times24\times24$ $\kv$-grid for bulk Mg. The band structures within the points have been determined as described in the computational details. The zoomed region magnify, respectively, for Na, the bandwidth (\textbf{b$_1$}) and the H point  (\textbf{b$_2$}), where multiple bands crossings occur, and for Mg the bandwidth of the upper (\textbf{d$_1$}) and lower (\textbf{d$_2$}) metallic bands. \label{fig:bulk_united}}
\end{figure*}

\subsection{W-av for bulk metals: the plasmon peak.}
\label{section:results_3d}

In this section, we present $GW$ results obtained with the W-av method for two prototypical bulk metals: Na, whose parabolic metallic band resembles the homogeneous electron gas (HEG) picture, and Mg, which exhibits a more complex band structure near the Fermi level with multiple metallic bands (the DFT bands are reported for reference as gray dashed lines in~\cref{fig:bulk_united}\textbf{b},\textbf{d}).

We begin by discussing the screening properties of these systems. In \cref{fig:bulk_united}\textbf{a},\textbf{c} we show the small-$\qv$ components of $\text{Im}[\overline{W}^{c}_{00}]$, obtained with and without the W-av method, in the region of the experimental plasmon that is represented in the plot as a Lorentzian function of 0.1~eV width (5.8~eV for Na~\cite{steiner1978kll,smith1969photoemission} and 10.35~eV for Mg~\cite{braun1983plasmon}). 
In the long-wavelength limit, the conventional calculation fails to reproduce the correct response of both materials, however in different ways. In the case of Na, the $\qv=0$ response is almost completely missing, with only a very broad and low-intensity structure appearing in the considered frequency range. In fact for bulk Na, similarly to  the HEG, the plasmon peak at $\qv=0$ stems only from intraband transitions that are not captured by the conventional scheme. The peak is, however, correctly reproduced within the W-av scheme. In Fig. S1 of the SI, we further show that the real part of the screening computed with W-av matches the Lindhard function of a homogeneous electron gas with $r_s = 3.94$. 
For Mg, on the other hand, the conventional calculation at $\qv = 0$ features a well-defined peak at an energy significantly lower than the experimental value. This peak originates from the interband part of the response, which is sizeable in bulk Mg. By including also the intraband contributions via the W-av method, the long-wavelength response is blue-shifted to the correct experimental energy. At finite $\qv$, both for bulk Na and Mg, the calculations with and without the W-av method give similar results, with only minor differences that decrease with increasing $\qv$. The curves at finite $\qv$ describe the dispersion of the bulk plasmon. 

~\cref{fig:bulk_united}\textbf{b} shows the DFT (grey) and $GW$ (orange) band structure of bulk Na. 
The magnitude of the $GW$ corrections is generally smaller than in the 2D case, becoming sizeable only at energies at least 2.5~eV away from the Fermi level. In particular, the corrections are largest at the extrema of the parabolic bands, e.g., at the bottom of the metallic band around $\Gamma$~(\cref{fig:bulk_united}\textbf{b1}), leading to a renormalization of the bandwidth by almost 300~meV, and near the H point~(\cref{fig:bulk_united}\textbf{b2}), where the corrections are also in the order of 250/300~meV. Figure~\ref{fig:bulk_united}\textbf{d} shows the DFT (grey) and $GW$ (orange) band structure of bulk Mg. Even in this case the $GW$ corrections are relatively small across the whole Brillouin zone, amounting to just few tens of meV within a 5~eV range from the Fermi level. Still, a renormalization of the bandwidths is observed, with the detailed zooms shown in~\cref{fig:bulk_united}\textbf{d1} and~\cref{fig:bulk_united}\textbf{d2}. For instance, the bandwidths of the two highest occupied states at $\Gamma$ are each changed by only about 50~meV relative to the DFT result.

\begin{table}[h]
    \centering
    \begin{tabular}{ccccc}
    \hline\hline \\[-6pt]
    & \multicolumn{2}{c}{This work} & 
       \multicolumn{2}{c}{Literature} \\
      \\[-8pt]
           &  PBE  & $G_0 W_0$  & $G_0 W_0$ & Exp.
           \\
\\[-10pt]
\midrule
\\[-6pt]
       Na & 3.23 & 2.94 &  2.85$^*$, 2.97$^{**}$$^{\textrm{~\citenum{Leon_2023}}}$, & 2.5$^{\textrm{~\citenum{jia_evidence_2022}}}$,\\ &&& 3.15$^{\dagger,\textrm{~\citenum{mandal_2022}}}$,
       3.04$^{\dagger,\textrm{~\citenum{friedrich_2022}}}$ &  2.65$^{\textrm{~\citenum{Lyo1988PRL}}}$  \\
\\[-6pt]
\midrule
\\[-6pt]
       Mg & 1.26 & 1.33 &  
       1.29$^{\dagger,\textrm{~\citenum{mandal_2022}}}$, 1.24$^{\dagger,\textrm{~\citenum{friedrich_2022}}}$ & 0.9$^{\textrm{~\citenum{bartynski1986}}},$ 1.24$^{\textrm{~\citenum{schiller2004}}}$ \\  
       & 1.70 & 1.65 &  
       1.68$^{\dagger,\textrm{~\citenum{mandal_2022}}}$, 1.63$^{\dagger,\textrm{~\citenum{friedrich_2022}}}$ & \hspace{-0.9cm} 1.79$^{\textrm{~\citenum{bartynski1986,schiller2004}}}$  \\  
       & 6.95 & 6.58 &  
       6.66$^{\dagger,\textrm{~\citenum{mandal_2022}}}$, 6.58$^{\dagger,\textrm{~\citenum{friedrich_2022}}}$ & 6.15$^{\textrm{~\citenum{bartynski1986}}}$, 6.89$^{\textrm{~\citenum{schiller2004}}}$   \\  [6pt]
           \hline\hline
    \end{tabular}
    \caption{Energy (in eV) of the last occupied electronic state at the $\Gamma$ point in bulk Na and bulk Mg. * Computed with PPA. ** Computed with MPA using a 16x16x16 k-grid. $\dagger$ Computed with contour deformation techniques, estimating the intraband plasmon with the tetrahedron method.}
    \label{tab:tab_bandwidth}
\end{table}

We have summarized our findings for the calculated bandwidths in~\cref{tab:tab_bandwidth} and we compare with data previously reported in literature. Our findings are in general in good agreement with previous $G_0 W_0$ calculations~\cite{mandal_2022,friedrich_2022} which treat metallicity by using dense $\kv$-grids as a base for the tetrahedron method, despite the different starting point that is PBE here and LDA in the previous works.
As anticipated, though excellent agreement with existing computational literature is found, the bandwidths calculated in this work differ significantly from the experimental counterparts in both bulk metals. As highlighted in previous studies~\cite{mandal_2022, friedrich_2022}, self-consistent $GW$ calculations (beyond the scope of this work) may be needed to improve the prediction of bulk-metals bandwidths.


\subsection{Acceleration of the convergence with respect to the \texorpdfstring{$\kv$}{k}-grid density}
\label{sec:3}
\begin{figure*}[t!]
    \centering  
     \centering   \includegraphics[width=1.01\textwidth,trim={0.0cm 0.0cm 0.0cm 0.0cm },clip]{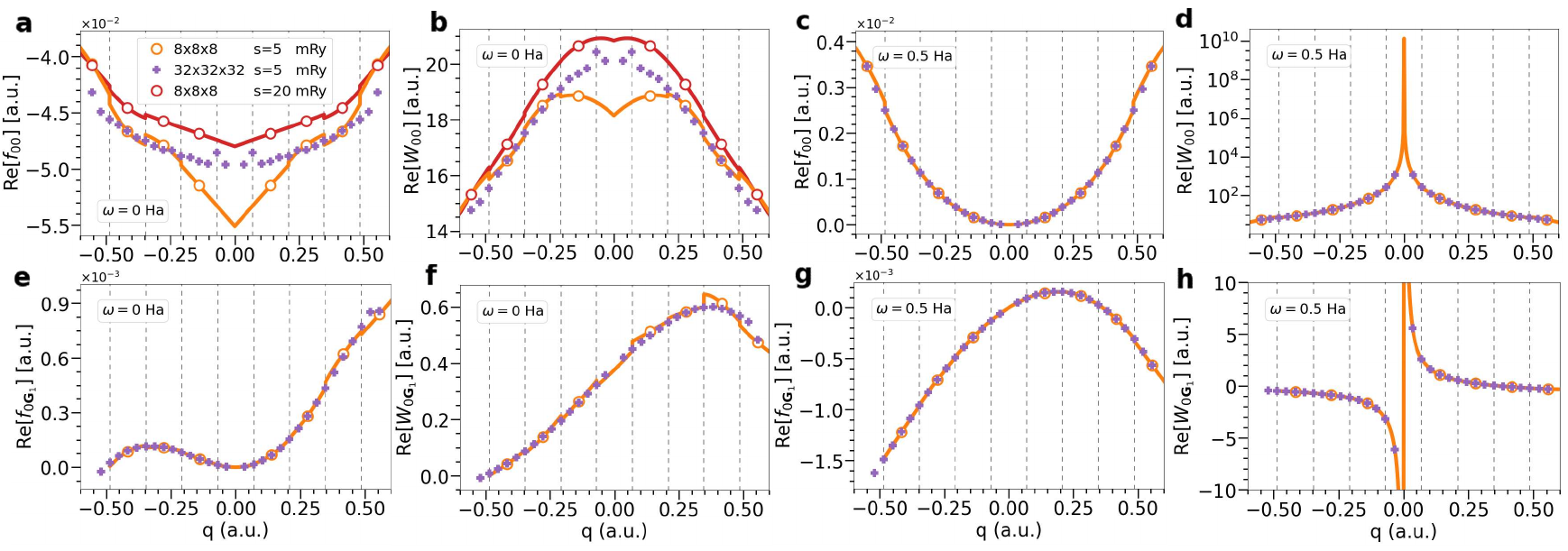}
    \caption{Result of W-av method in bulk Na on the head auxiliary functions $f_{0 0}$ (panels \textbf{a},\textbf{c}) and wing $f_{0 \mathbf{G_1}}$ (panels \textbf{e, g)}, at zero and finite frequency. The wing element has $\mathbf{G_1}=(0,-0.5,-0.5)$. The corresponding screened potential $W_{0 0}$ is shown in panels \textbf{b}, \textbf{d} and $W_{0 \mathbf{G_1}}$ in panels \textbf{f}, \textbf{h}. Orange (violet) dots are calculated with the $8\times 8\times 8$ ($32\times 32\times 32$) grid. Continuous orange and red lines are obtained with the W-av method starting from the $8\times 8\times 8$ grid for different smearing values $s$ (5 and 20 mRy).  \label{fig:interp_head_3D}}
\end{figure*}
\begin{figure*}[t!]
    \centering  
      \includegraphics[width=\textwidth,clip]{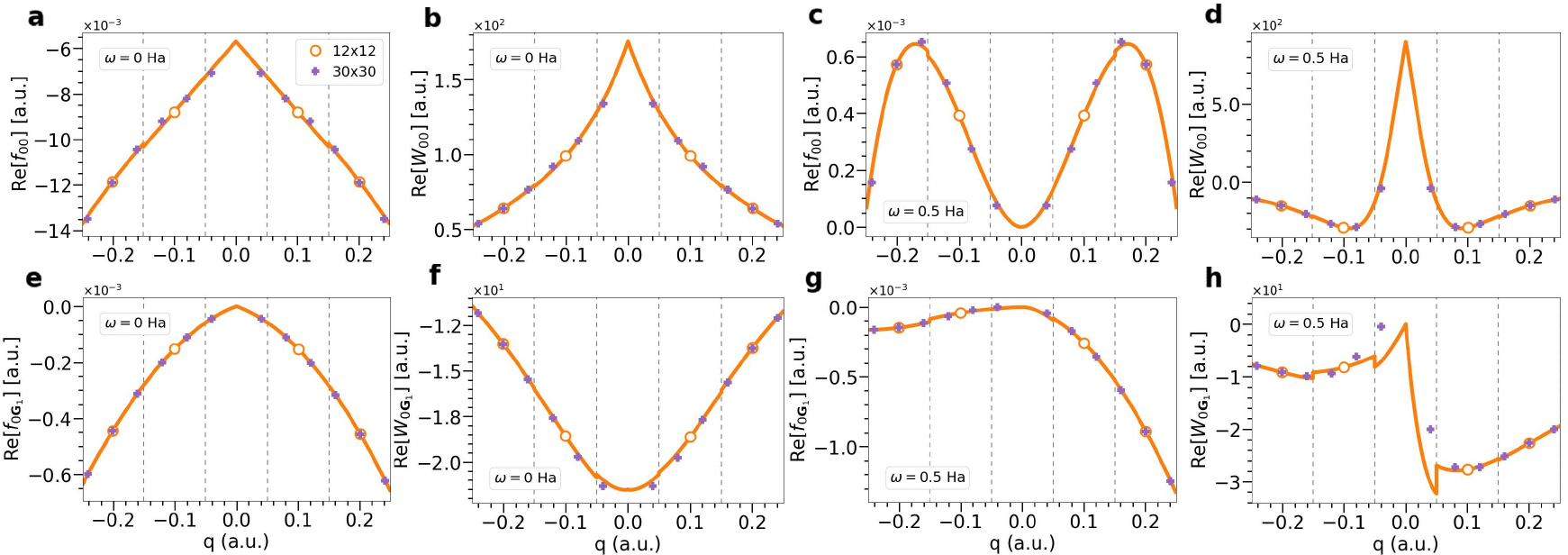}
    \caption{
    Results of W-av method in $n$ doped monolayer MoS$_2$ ($\rho = 0.04$ el/cell) on the head auxiliary functions $f_{0 0}$ (panels \textbf{a}, \textbf{c}) and wing $f_{0 \mathbf{G_1}}$ (panels \textbf{e}, \textbf{g}), at zero and finite frequency ($\omega=0.5$ Ha). The wing element has $\mathbf{G_1}=(0,0,1)$. The corresponding screened potential $W_{0 0}$ is shown in panels \textbf{b}, \textbf{d} and $W_{0 \mathbf{G_1}}$ in panels \textbf{f}, \textbf{h}. Orange (violet) dots are calculated with the $12\times 12$ ($30\times 30$) grid. Continuous orange lines are obtained with the Wav interpolation over starting from the $12\times 12$ grid.
    \label{fig:interp_2D}}
\end{figure*}

In this Section we first perform a detailed analyses of the W-av analytic reconstruction of the screened potential, used in the random integration of~\cref{Eq_W_av}. Next, we discuss the speed-up achieved in $GW$ calculations with respect to the $\kv$-point sampling. In this analysis, we also show results with the previously developed constant approximation (CA)~\cite{Leon_2023}, which consists in approximating the long-wavelength limit of $\epsilon^{-1}$ with its value calculated at the smallest available $\qv_{min}\neq0$, i.e.,  $\epsilon^{-1}_{\mathbf{G},\mathbf{G'}}(\qv\to0,\omega)=\epsilon^{-1}_{\mathbf{G},\mathbf{G'}}(\qv_{min}, \omega)$. In practice, this is analogous to an extrapolation to $\qv\to0$, using a constant function. Finally, we will discuss the methodological challenges addressed here in the case of graphene.

In~\cref{fig:interp_head_3D,fig:interp_2D}, we show for a coarse (empty orange dots) and a dense $\kv$-grid (purple dots) the calculated values of $W$ and the auxiliary functions $f$ used for the extrapolation/interpolation. We compare with the correspondent analytic forms reconstructed by the W-av procedure on the coarse $\kv$-grid. 
We recall that the long-wavelength analytic behavior of $f$ and $W$ is given by the expressions in~\cref{table:q3D,table:q2D}. The orange solid lines correspond to the head and wing terms of $f$ and $W$ for a 3D case (bulk Na) and a 2D case (MoS$_2$ with a $n$ doping of 0.04~el/cell), respectively. The static and the finite-frequency cases are shown in separate panels. \cref{fig:interp_head_3D}\textbf{a},\textbf{b} also display red line and dots, these are results obtained with a different smearing that are discussed in detail later.

For bulk Na, the W-av method applied to the coarse $8 \times 8 \times 8$ grid successfully reconstructs the missing long-wavelength response, closely matching the results from the denser $32 \times 32 \times 32$ grid. There is, though, a minor discrepancy in the static head term, that has not yet fully converged on the $8 \times 8 \times 8$ grid (see \cref{fig:interp_head_3D} \textbf{a},\textbf{b}). Nevertheless, this does not pose a problem for QPs calculations, as shown in section \ref{section:results_3d}, since the dynamical part of the response, which provides the most significant contribution, has reached convergence on the $8 \times 8 \times 8$ grid (see \cref{fig:interp_head_3D} \textbf{c},\textbf{d}).

The slower convergence of the static response with respect to the k-point mesh compared to that of the QP corrections arises from its dependence on the occupations near the Fermi level, which are influenced by the $\kv$-space discretization and the chosen electronic distribution (in this case, a Fermi-Dirac distribution with a width of 5 mRy). 
To explore this further, we have performed $GW$ calculations with the W-av method on the $8 \times 8 \times 8$ grid using different Fermi-Dirac smearings: 5 mRy (orange curve) and 20 mRy (red curves \cref{fig:interp_head_3D} \textbf{a},\textbf{b}). In the long-wavelength regime of the static case, the smaller smearing value only approximates the Yukawa behavior, which instead is observed with the denser $32 \times 32 \times 32$ grid. Increasing the smearing to 20 mRy significantly improves the agreement between the coarse and dense grids, indicating that a larger smearing smooths the $\chi V$ at small $\qv$, compensating for the larger grid spacing and improving the extrapolation. However, since the static head term has a small weight in the response, the QP energies change only slightly within reasonable smearing values, as shown in Fig. S2 of the SI.

Also in the 2D case, the W-av procedure successfully reconstructs the missing long-wavelength limit, as shown in~\cref{fig:interp_2D} for doped monolayer MoS$_2$. For the terms shown, the result of the W-av procedure on the coarse grid is similar to the numerical results from the denser grid computed without W-av. We observe only minor differences in the static term, likely due again to the sensitivity of this term to the Fermi surface. For the dynamical terms, W-av accurately captures the long-wavelength limit even for terms of $W$ where the coarse-grid behavior is nontrivial, such as for the wing term of panels (\textbf{g}) and (\textbf{h}).

\begin{figure}[t]
    \centering 
    \includegraphics[width=0.45\textwidth,trim={0.5cm 0.2cm 0 0.0cm },clip]{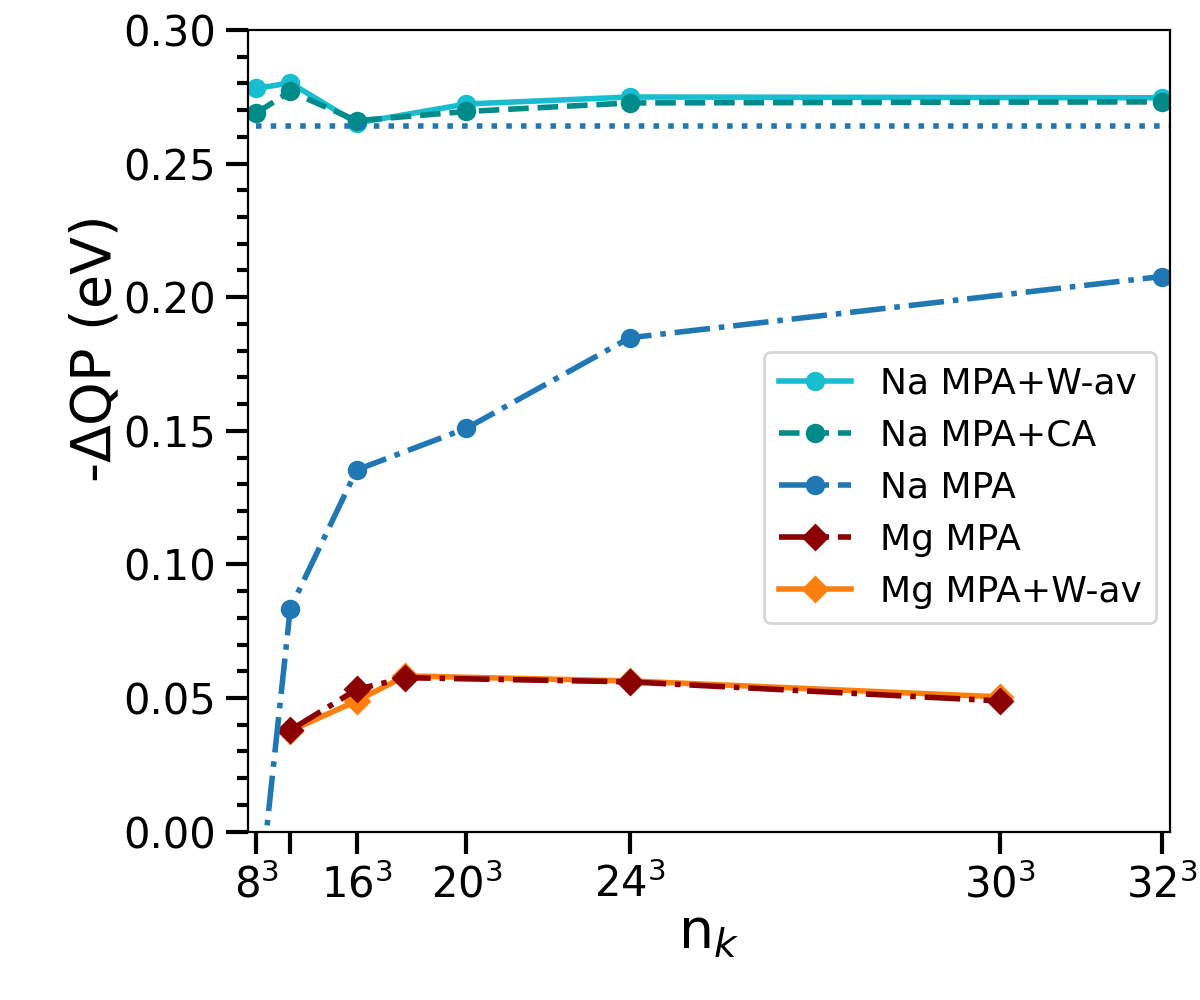}
    \caption{
    $GW$ QP correction to the bandwidths of 3.23~eV of bulk Na and of 1.70~eV of bulk Mg, as a function of the number of points in the $\kv$-grid, computed with (continuous line) and without (dash-dotted line) the W-av method. In the case of bulk Na, we also compare with the constant approximation (CA, dashed line)~\cite{Leon_2023}. The dotted blue line is the extrapolation to an infinite $\kv$-grid of a standard MPA calculation for bulk Na; the extrapolated data lies 11 meV apart from the result obtained using the W-av method  with the most dense grid (though convergence is achieved rapidly and the QP difference is almost constant over the range of grids).}
    \label{fig:bandwidth_Na}
\end{figure}

In the following, we analyze the $\kv$-grid convergence for the materials studied. We first consider two selected $GW$ bandwidths for bulk Na and Mg, previously discussed in Section~\nameref{section:results_3d}. To study the convergence we did not perform QP calculations of the full BZ for all the considered $\kv$-grid thus the value of $\Sigma$ at Fermi in~\cref{eq.gw_ex} is determined by considering the $GW$ corrections for two points close to the Fermi level and assuming a linear variation of $\Sigma$ between them. The resulting value of $\Sigma$ was found to be consistent with that resulting from the full BZ calculations of the metallic bands in selected cases.

The corresponding $GW$ results are shown in~\cref{fig:bandwidth_Na}, along with the extrapolated values on an infinitely dense $\kv$-point grid. The extrapolation is performed using a function of the form $f(n_k) = a-b/\sqrt{n_k}-c/n_k$, where $n_k$ is the number of $\kv$-points in the BZ. This expression has been shown~\cite{Marrazzo_2019,Bonacci2023npjComputMater} to describe well the convergence behavior fo QP corrections with an increasing number of $\kv$-points, allowing one to extrapolate the QP corrections to an infinite $\kv$ grid. 
Although as discussed earlier the conventional approach fails to reproduce the long-wavelength frequency response for both Na and Mg (see ~\cref{fig:bulk_united}\textbf{a}-\textbf{c}), the convergence behavior of the QP corrections differs markedly between the two materials. For Mg, convergence within 20 meV is already achieved with a coarse $12\times12\times12$ grid, independently of the treatment of long-wavelength intra-band contributions. This is due to the strong inter-band response in the $\qv \to0$ limit. Despite the incorrect energy position of its plasmon, as the QP energies lie far from the structures of $\Sigma$ originating from the plasmons, an approximate description is sufficient to correctly evaluate the low-energy corrections. 
In contrast, the inter-band response in bulk Na is negligible, making the inclusion of the missing intra-band contribution essential. As a result,  QP calculations done without proper treatments converge very slowly with respect to $\kv$-point sampling: even the finest $\kv$-grid used ($32\times32\times32$) differs by about 60~meV from the extrapolated value of 0.264~eV. When the long-wavelength intraband contribution is included, either via the W-av method or the CA, convergence within 11~meV is achieved already with an $8\times8\times8$ grid. The agreement between the W-av method and CA arise from the slow (almost constant) $\sim q^2$ variation of the dielectric matrix close to $\qv=0$, making the CA a reliable methodology for bulk metals.

From this two cases, we expect the W-av treatment of inter-band response to be important for any bulk metal with a $\Sigma$ that presents structures close to the QP energies, and negligible in the cases when the structures of $\Sigma$ lie at high energies. As described in the following, this is different for 2D metals where the metallic plasmon occurs at low energies, close to the QP solution, and the plasmon energy varies sharply close to $\qv\to0$.

\begin{figure}[t]
   \centering   \includegraphics[width=0.43\textwidth, trim={0.1cm 0.4cm 0.1cm 0.0cm },clip]{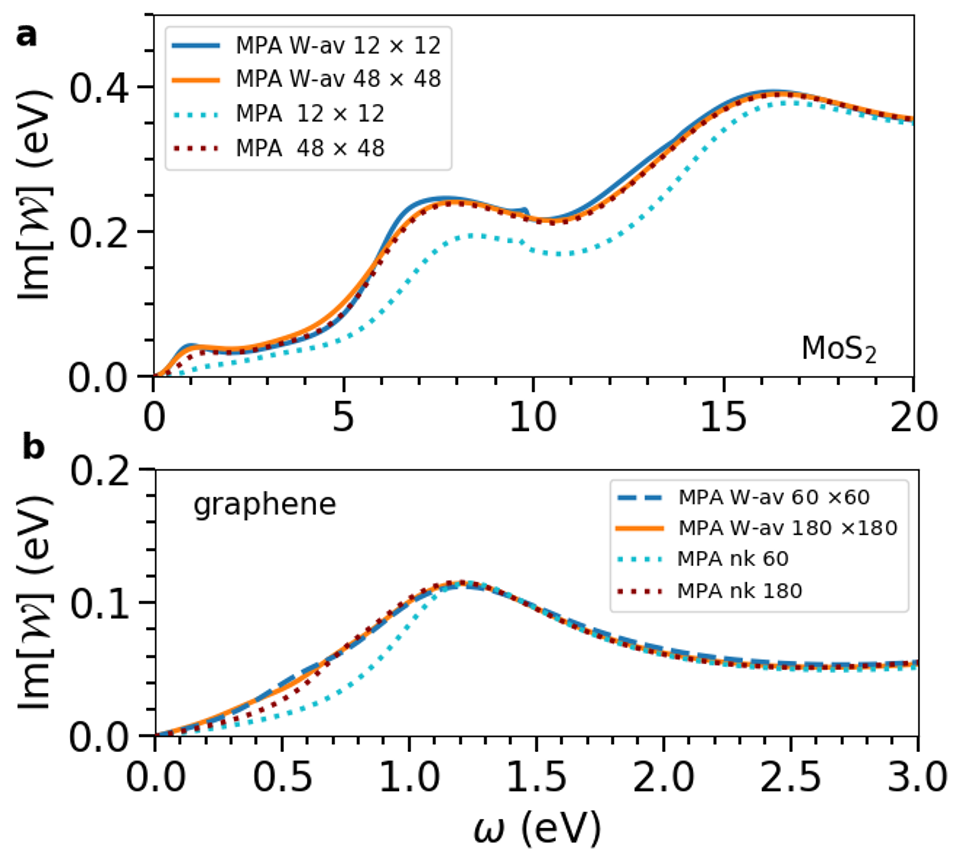} 
    \caption{Imaginary part of the screened Coulomb potential $\mathcal{W}(\omega)$: a) MoS$_2$ $n$ doped with $\rho$=4.56$\times 10^{13}$ el/cm$^{2}$ on calculations on a $12\times12$ and a $48\times48$ $\kv$-grid with and without using the W-av method;  b) Graphene with a $p$ doping of -0.1 el/cell on calculations on a $60\times60$ and a $180\times180$ grid with and without using the W-av method. In plot~\textbf{b}, we only show the low-frequency response. \label{fig:comparison_W_2D}}
\end{figure}
%


In~\cref{fig:comparison_W_2D}\textbf{a}, we show the imaginary part of $\mathcal{W}(\omega)$ for doped MoS$_2$, as a function of the $\kv$-grid size. Without W-av, the standard calculation on a $12\times12$ $\kv$-grid severely underestimates both the metallic 2D plasmon and the inter-band plasmon around 8~eV. With a $48\times48$ $\kv$-grid, the result is much closer to that obtained with W-av, although the low-energy metallic plasmon is still not fully converged. Only with W-av the integrated Coulomb potential $\mathcal{W}(\omega)$ is similar for the coarse and dense $\kv$-grids, demonstrating the efficiency of the W-av procedure. The $GW$ calculations for 2D systems are typically performed with much denser grids~\cite{Liang2015,Gao2017}.

\begin{figure*}[t!]
    \centering 
     \centering  \includegraphics[width=\textwidth,trim={0.0 0.4cm 0.0cm 0.2cm},clip]{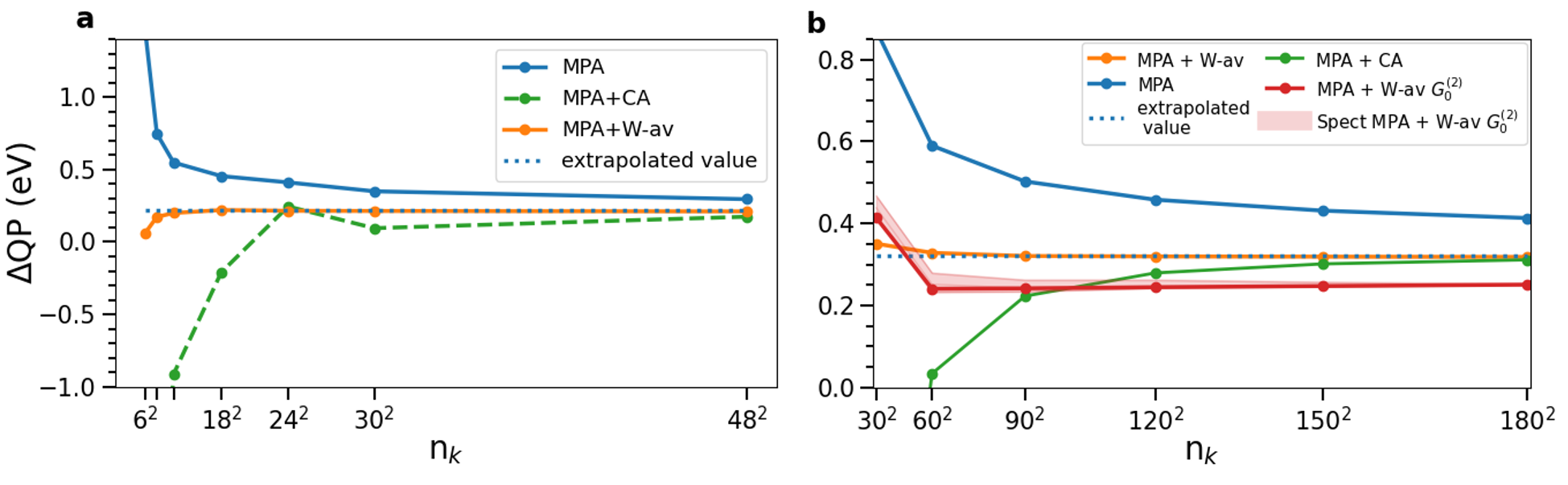} 
    \caption{
    Convergence of the QP corrections with respect to the $\kv$-grid for \textbf{a)} the K point band gap of MoS$_2$ $n$ doped with 0.04 el/cell and \textbf{b)} the M point band gap of doped graphene with a p doping of -0.1 el/cell. We show the result of the G$_0$W$_0$ calculation with the MPA (blue line), MPA + CA (green line) and MPA + W-av (orange line). The blue dots have been obtained extrapolating the MPA calculation on an infinite $\kv$-grid. The \textbf{b)} plot also shows a red colored region, this indicates the QP corrections determined by the spectral solution using an higher order propagator G$^{(2)}_0$ using a range of damping between 0.15 and 0.3~eV. The spectral solution can be approximated by red line, the linearised solution of a MPA + W-av calculation performed with the second order propagator and damping of 0.2~eV.} \label{fig:bandwidth_conv_2D}
\end{figure*}

In ~\cref{fig:bandwidth_conv_2D}\textbf{a}, we compare the energy difference between the first band above and below the Fermi level, at the $K$ point of doped MoS$_2$, calculated using various $\kv$ grids and different treatments of the $\qv\rightarrow 0$ limit. 
If computed without any long-wavelength correction, the quasi-particle band separation converges slowly, approaching convergence only with a $48\times48$ grid. The CA correction to the long-wavelength behavior, shown above to be successful for 3D systems, is much less effective in the 2D case. CA treats the correlation part of the dielectric matrix $\epsilon_{\mathbf{G},\mathbf{G}'}-1$ as a constant in the small $\qv$ limit. While this is a  reasonable approximation for 3D systems, it breaks down for 2D metals. Therefore, the CA reproduces the correct $\qv\rightarrow 0$ divergence of the correlation part of the screened Coulomb potential in the bulk case, \cref{fig:interp_head_3D}\textbf{d}, but fails in the 2D case, where instead $W^c$ approaches a constant with a sharp variation (see, e.g.,~\cref{fig:interp_2D}\textbf{d}).
In contrast, with W-av, the QP corrections already match the extrapolated value on a $12\times12$ $\kv$-grid, thereby strongly accelerating the $\kv$-space convergence.

We finally discuss the case of graphene.
Calculations of pristine graphene are known to be significantly more demanding than those of 2D semiconductors, reaching convergence only for very dense $\kv$-grids. While a $18\times18$ grid is sufficient for converged DFT results, $GW$ calculations require a $180\times180$ grid without W-av~\cite{GuandaliniLeon_2024}, that reduces to a $60\times60$ grid with W-av. The need for dense $\kv$-grid arises from the large number of points required in the Dirac-cone region to describe the RPA response of~\cref{eq.polarizability}.
In doped graphene, although the Fermi level does not coincide with the Dirac point, the Fermi surfaces are small and the Dirac cones have the same steep dispersion. Even if the W-av method reduces the required density of the $\kv$-grids, they will still be denser than for other 2D systems.
Furthermore, in 2D metals, the frequency of the intra-band peak ($\omega_p$) depends on the momentum and approaches zero following $\omega_p\sim\sqrt{q}$~\cite{hwang2007dielectric}. As a result, at small $\qv$,
$\overline{W}^{c}(\qv,\omega)$ is given by a sequence of multiple peaks (see Fig. S3 in Section S IV of the Supplementary Information, SI). Making the $\kv$-grids denser, the number of peaks of $\overline{W}^{c}(\qv,\omega)$ increases and at the same time the spacing reduces, up to the infinitely dense $\kv$-grid limit where a broad intra-band peak is established. It is possible to obtain a smooth intra-band plasmon from the integration of $\overline{W}^{c}(\qv,\omega)$ at reasonable $\kv$-grid sizes introducing a suitable Lorentzian broadening, with coarser grids requiring larger broadening values.

In the calculation of the self-energy, as defined in~\cref{eq_GW_expl}, the discrete integration over momentum $\qv$ is weighed by the one particle Green's function $G_0$, given by the sum of simple poles with a Lorentzian broadening. 
The tails of the Lorentzian functions for different $\qv$ values, decay slowly (1/$\omega^2$). Their overlap results in a $\Sigma$ that depends on the broadening. As explained in detail in Section S IV, it is possible to overcome this dependence by combining the large broadening used to evaluate $W$ with the use of 2$^{nd}$-order generalized Lorentzian poles in the description of the Green function~\cite{chiarotti2022unified}, where $\text{Im}[G^{(2)}]\propto
\mathcal{L}_j^{(2)}(\omega)$ and
\begin{eqnarray}
    \label{eq:higher_order_GF}
    \mathcal{L}_j^{(2)}(\omega)= \frac{\sqrt{2}}{\pi} \frac{\eta^3}{(\omega-\epsilon_j)^4+ \eta^4}.
  \end{eqnarray}
The tails of the imaginary parts of the 2$^{nd}$-order generalized Lorentzian decay faster than the standard Lorentzian functions, $1/\omega^4$ instead of $1/\omega^2$, reducing their overlap. In the following paragraphs, we discuss results obtained using both first- and second-order (here referred as $G_0^{(2)}(\omega)$) Lorentzian Green’s functions, showing that the latter provide more robust numerical results.

We first analyze the dependence of $\mathcal{W}(\omega)$ on the $\kv$-grids in $p$-doped graphene, considering two grid sizes: $60\times60$ and $180\times180$. The low-energy behavior of $\mathcal{W}(\omega)$, shown in~\cref{fig:comparison_W_2D}\textbf{b}, is compared with and without the W-av method. As discussed in Section~\nameref{section:bandgap_renorm_2D} when presenting the case of doped graphene, doping affects primarily this region that features the intra-band (Dirac) plasmon. In the summation for $\mathcal{W}(\omega)$ we apply a Lorentzian broadening of 0.2~eV, larger than that used for MoS$_2$, for the reasons outlined above and further discussed i SEction S IV of the SI. Without W-av, only the dense 180$\times$180 $\kv$-grid reproduces the expected behavior of the intra-band plasmon. By contrast, with W-av the intra-band plasmon is already converged with the coarser 60$\times$60 $\kv$-grid.

Next, we examine the convergence of the QP energies, for which the accuracy of the Green's function plays a fundamental role. In~\cref{fig:bandwidth_conv_2D}\textbf{b}, we show the convergence of the band gap at the $M$-point of $p$-doped (-0.1~el/cell) graphene as a function of the number of $\kv$-points. The results from the standard MPA calculations, using $G^{(1)}(\omega)$, and the corresponding extrapolation are shown by solid and dashed blue lines, respectively. The green and orange curves include intraband corrections in the MPA calculations using CA and W-av. 
As seen for MoS$_2$, QP corrections of a standard MPA calculation, shown in \cref{fig:bandwidth_conv_2D}\textbf{a}, converge slowly. The discrepancies with respect to the extrapolated value are still larger than 50~meV even with a $180\times180$ grid. The CA accelerates convergence but is less effective than in bulk systems, approaching the extrapolated value only for grids larger than $120\times120$. In contrast, the W-av method  reproduces the extrapolated QP correction already with the $60\times60$ grid.

Due to the difficult handling of the intraband plasmon discussed earlier, the  QP corrections show a strong dependence on the broadening $\eta$ and, in particular, do not coincide with the corrections determined from the maximum of the spectral function.
For this reason, we also report in red in~\cref{fig:bandwidth_conv_2D}\textbf{b} QPs calculated with W-av but using the 2$^{nd}$-order Lorentzian $G_0^{(2)}(\omega)$.
The shaded red area corresponds to solutions obtained from the peaks of the spectral function~\cref{eq.spectralfunc}, encompassing dampings from 0.1 to 0.3~eV. These spectral solutions calculated with $G_{0}^{(2)}$ are found to be very robust with respect to the damping. This is further illustrated in Fig. S5 in the SI where we compare them with the spectral solutions with $G_0^{(1)}(\omega)$. The spectral solutions also exhibit a rapid convergence with the $\kv$-grid, but they approach a slightly different asymptotic value with respect to the result of the W-av method with $G_0^{(1)}$ in blue. Moreover, at dense $\kv$-grids, the shaded area narrows as the dependence on the damping decreases.

However, when a large number of QP corrections has to be calculated, plotting the spectral function for each of the QPs becomes unpractical. In such cases, it is desirable to employ the linearized QP equation. By comparing linearized solutions obtained with different broadening parameters to the QP energies extracted from the peaks of the spectral function, for few QP corrections, it is possible to determine an optimal broadening parameter that yields consistent results between the two approaches. In this case, the linearized QP solutions, obtained with a second-order propagator $G_{0}^{(2)}$ and a damping parameter of 0.2 eV (solid red line in~\cref{fig:bandwidth_conv_2D}\textbf{b}) reproduce the solutions given by the peaks of the spectral function.

The agreement between the two approaches is consistent across all the doping levels studied, as determined from calculations on the $60\times60$ $\kv$-grid (Fig. S6 of the SI). 
We also find that the differences between the solutions obtained using $G_{0}^{(1)}$ and $G_{0}^{(2)}$ are not constant but increase with doping. 

\section{Discussion}
\label{section:conclusions}
%
In this work, we have addressed GW calculations for metallic systems by introducing a fully ab initio approach, W-av, which explicitly accounts for intraband contributions in the evaluation of the response function for both 3D and 2D metals, without relying on external parameters. The W-av method, by handling separately long-range intraband contributions in the static and dynamic limit, is well suited to be applied in combination with analytical frequency descriptions for $W$, such as plasmon-pole or multipole models. 
We have demonstrated that W-av in combination with MPA enables an accurate description of doping-induced effects, as illustrated by the gap renormalization in monolayer MoS$_2$ and graphene, as well as the Drude peak in simple metal as Na and Mg.
A key advantage of W-av is the substantial improvement in the convergence of GW calculations with respect to $\kv$-point sampling. This effect is especially pronounced in 2D systems, where conventional approaches ---even when employing very dense $\kv$-grids--- can yield quasiparticle corrections that remain hundreds of meV away from the converged limit.
By contrast, W-av achieves rapid convergence, significantly reducing the computational cost associated with dense Brillouin zone sampling. 

Importantly, the additional computational overhead introduced by the method is negligible when compared to the gains of using reduced k point grids. 
The improved convergence originates from the combined treatment of the screened interaction: on the one hand, through interpolation in discretized $\qv$-space~\cite{Guandalini2023npjCM}, and on the other, through a physically motivated extrapolation to the $\qv\rightarrow0$ limit that properly captures the intraband contribution to the irreducible polarizability. 
Overall, W-av provides a reliable and efficient framework for GW studies of metallic systems, particularly in situations where intraband effects and k-point convergence are critical. 

We also propose the use of higher order Lorentzians~\cite{chiarotti2022unified} in the description of the self-energy and of a full-frequency solution of the QP equation beyond linearization. This new scheme has shown to be relevant for graphene, that presents a marked dependence of the QP corrections on the damping parameter used in the self-energy calculation.\\

\section{Methods}
\subsection{Computational details}
\label{sec:methods}

In the present work we investigate both 3D and 2D metallic systems: bulk Na and bulk Mg, 2D MoS$_2$ and graphene with various levels of doping. The crystal structures of all systems were relaxed at the DFT level, minimizing the total energy with respect to ionic coordinates and lattice parameters. The DFT calculations were performed using the PBE~\cite{Perdew1996} exchange and correlation functional, except for graphene, for which we used LDA~\cite{Perdew-Zunger1981PRB}, using norm-conserving pseudo-potential, as implemented in the \textsc{Quantum ESPRESSO} suite~\cite{Giannozzi2009JPCM_v2,Giannozzi2017JPCM_v2}. Many body perturbation theory calculations such as $GW$ are instead performed using the \textsc{Yambo} code~\cite{yambo_2019}. The GW calculations are performed with a damping for G of $\eta=0.1~eV$, unless explicitly indicated.

According to our simulations, 
Bulk Na crystallizes in a BCC lattice with a computed lattice constant of $a=4.23$~\AA. The DFT calculations were performed using  a Fermi-Dirac smearing of 0.005 Ry and a wavefunction (WF) cut-off of 100 Ry.  $GW$ calculations were performed using the recently developed MPA method for the matrix $\chi$, with 8 frequency poles chosen with a linear sampling between 0 and 4~Ha, 100 bands, and a cutoff of 20 Ry. The self-energy, $\Sigma$, was respectively calculated using 100 bands for the study of the $k$-grid convergence and 500 bands for QPs calculations.
For the HCP structure of Mg, the computed lattice constants are $a = 3.193$~\AA\ and $c = 5.188$~\AA. Calculations adopted a WF cut-off of 120~Ry and a Methfessel-Paxton (MP) smearing of 0.01~Ry. In $\yambo$, the occupations are described with a Fermi-Dirac smearing of 0.038~eV that allows one to match the DFT Fermi energy.
The $GW$ simulations are carried out using the MPA approach with 8 frequency poles chosen with a linear sampling between 0 and 4~Ha and 240 bands in the polarizability, together with a 
$\chi$ matrix cutoff of 20 Ry; the $GW$ $\Sigma$ was calculated including 500 bands.

Concerning 2D systems, we computed the relaxed lattice structure of MoS$_2$ and graphene in the undoped case, and found values of the lattice constants 3.184~\AA \ and 2.46~\AA, respectively. The insertion of free charges upon doping effectively changes the 2D lattice, as discussed in detail in the literature, particularly in the case of graphene~\cite{Lazzeri2006PhysRevLett}. 
However, in this work we neglect this effect since it goes beyond the scope of the present study. For the same reason, we do not consider the spin-orbit interaction, which would be needed for an accurate characterization of TMDs such as MoS$_2$.
Since all calculations are performed using periodic boundary conditions, in order to avoid interaction between replicas, the distance between layers was set to 9.84~\AA\ for graphene and to 10~\AA\ for MoS$_2$. The undoped calculation for MoS$_2$ was done with a vacuum of 20~\AA. Such spacing also prevents the formation of bound vacuum states, a well-known issue in DFT calculations in the presence of explicit carrier doping.

The DFT calculation for MoS$_2$ was done with a  Fermi-Dirac smearing of 0.005~Ry, and a cutoff of 60~Ry. The MBPT calculations at the $GW$ level were performed using the MPA method for the $\chi$ matrix, with 8 frequency poles chosen with a quadratic sampling between 0 and 4~Ha, 400 bands, and a WF cutoff of 10~Ry, whereas $\Sigma$ was calculated using 600 bands.
In the case of Graphene, the DFT computation was performed with a MP smearing of 0.01 Ry, and a WF cutoff of 90 Ry. In this case, the Fermi level determined by a MP smearing of 0.01~Ry is matched by a Fermi-Dirac smearing of 0.0325 eV. The MBPT calculations at $GW$ level were performed using the MPA method for the $\chi$ matrix, with 10 frequency poles chosen with a quadratic sampling between 0 and 4~Ha, 100 bands, and a cutoff of 5~Ry, whereas $\Sigma$ was calculated using 100 bands. Such a choice of the $GW$ parameters is sufficiently converged to accurately characterize the electronic band structure in the region of the graphene M points, that is studied in detail.

The study of the $GW$-calculations convergence with respect to the number of empty bands is shown in Section S V of the SI for all the studied materials. Our results are converged with respect to the number of empty states within a~10-15 meV range, upon including the QP correction at Fermi.

Concerning the plots of the band structures of ~\cref{fig:MoS2_united,fig:bulk_united}, the DFT bands reported have been determined from dense $nscf$ calculations. The correspondent $GW$ band structure have been calculated from the DFT result adding the interpolated $GW$ corrections determined using the SKW tool~\cite{gonze2020abinit} available in $\yambopy$~\cite{paleari2025yambopy}. As basis for the $GW$ interpolation, we have used the $k$-grid indicated in the captions of~\cref{fig:MoS2_united,fig:bulk_united}. In the case of~\cref{fig:graph_united}, instead, both the DFT and the $GW$ band structure result from an interpolation on top of the indicated $k$-grid. Also in this case, the interpolation is performed using the SKW tool available in $\yambopy$.

\section{Data availability}
The data that support the findings of this article are not publicly available upon publication because it is not technically feasible and/or the cost of preparing, depositing, and hosting the data would be prohibitive within the terms of this research project. The data are available from the authors upon reasonable request.

\section{Code availability}
The code developed in this work is available in the public git repository of the Yambo code.

\section{Acknowledgments}
\label{section:aknowl}
%
We acknowledge enlightening scientific discussions with D. Le\'on Valido, F. Paleari, and E. Molinari. We also thank N. Spallanzani for code-support, in particular with the GPU porting.
Partial support has been provided by Italian MUR through PRIN 2022 "TUNES" (Grant No. 2022NXLTYN), (GS, AF) and  European Union –NextGenerationEU,
through the MUR -- Prin 2022 programme (2D-FRONTIERS, Grant No. 20228879FT); 
by ICSC--Centro Nazionale di Ricerca in High Performance Computing, Big Data and Quantum Computing--funded by the European Union through the Italian Ministry of University and Research under PNRR M4C2I1.4 (Grant No. CN00000013) (AF, DV).
We acknowledge ISCRA and ICSC for awarding access to the LEONARDO supercomputer, owned by the EuroHPC Joint Undertaking, hosted by CINECA (Italy). 
This work has also been supported by the European Union through the MaX "MAterials design at the eXascale" Centre of Excellence (Grant agreement No. 101093374, co-funded by the EuroHPC JU) (AF, DV, PDA, CC).

\section{Author contributions}
GS devised and implemented the method, and performed the calculations. AG provided scientific and technical support. AF, PDA, CC, MR, and DV conceived the original idea of the work. GS and CC made major contributions to the manuscript writing. All authors contributed to the method development, analysis of the results, manuscript writing, and critically discussed the paper.

\section{Competing interests}

The authors declare no competing financial or non-financial interests

\renewcommand{\emph}{\textit}
\bibliography{biblio}

@misc{note_yambo_vars,
  note={In \textsc{Yambo}, the procedure is activated by the user through a single flag in the input file, 
  \texttt{rimw\_type= "metal"}},
  year = {},
}

@ARTICLE{Kohn1965PhysRev,
 author = {Kohn, W. and Sham, L. J.},
 issn = {0031-899X},
 journal = {Phys. Rev.},
 month = {nov},
 number = {4A},
 pages = {A1133–A1138},
 publisher = {American Physical Society (APS)},
 title = {Self-Consistent Equations Including Exchange and Correlation Effects},
 doi = {http://dx.doi.org/10.1103/PhysRev.140.A1133},
 volume = {140},
 year = {1965}
}

@ARTICLE{ismail2006truncation,
  title={Truncation of periodic image interactions for confined systems},
  author={Ismail-Beigi, Sohrab},
  journal={Phys. Rev. B—Condensed Matter and Materials Physics},
  volume={73},
  number={23},
  pages={233103},
  year={2006},
  doi={https://doi.org/10.1103/PhysRevB.73.233103},
  publisher={APS}
}

@ARTICLE{Engel1993PRB,
   author = {G.~E. Engel and Behnam Farid},
   title = {Generalized plasmon-pole model and plasmon band structures of crystals},
   journal = {Phys. Rev. B},
   volume = {47},
   number={23},
   pages = {15931},
   year = {1993},
   doi = {https://doi.org/10.1103/PhysRevB.47.15931}
}

@article{Engel1992PRB,
 author = {Engel, G. E. and Farid, Behnam},
 doi = {10.1103/physrevb.46.15812},
 issn = {1095-3795},
 journal = {Phys. Rev. B},
 month = {Dec},
 number = {24},
 pages = {15812–15827},
 publisher = {American Physical Society (APS)},
 title = {Calculation of the dielectric properties of semiconductors},
 url = {http://dx.doi.org/10.1103/PhysRevB.46.15812},
 volume = {46},
 year = {1992}
}

@ARTICLE{Hybertsen1986PRB,
   author = {Mark~S. Hybertsen and Steven~G. Louie},
   title = {Electron correlation in semiconductors and insulators: Band gaps and quasiparticle energies},
   journal = {Phys. Rev. B},
   volume = {34},
   number = {},
   pages = {5390},
   year = {1986},
   doi = {https://doi.org/10.1103/PhysRevB.34.5390}
}

@ARTICLE{Godby1989PRL,
   author = {R.~W. Godby and R.~J. Needs},
   title = {Metal-insulator transition in Kohn-Sham theory and quasiparticle theory},
   journal = {Phys. Rev. Lett.},
   volume = {62},
   number = {},
   pages = {1169},
   year = {1989},
   doi = {https://doi.org/10.1103/PhysRevLett.62.1169}
}

@ARTICLE{Marini2001PRL,
   author = {Andrea Marini and Giovanni Onida and Rodolfo Del~Sole},
   title = {Quasiparticle Electronic Structure of Copper in the \textit{GW} Approximation},
   journal = {Phys. Rev. Lett.},
   volume = {88},
   number = {1},
   pages = {016403},
   year = {2001},
   doi = {https://doi.org/10.1103/physrevlett.88.016403}
}

@ARTICLE{Golze2019FrontChem,
   author = {Dorothea Golze and Marc Dvorak and Patrick Rinke},
   title = {The \textit{GW} {C}ompendium: {A} {P}ractical {G}uide to {T}heoretical {P}hotoemission {S}pectroscopy},
   journal = {Front. Chem.},
   volume = {7},
   number = {},
   pages = {377},
   year = {2019},
   doi = {https://doi.org/10.3389/fchem.2019.00377}
}

@ARTICLE{Aryasetiawan1998RPP,
   author = {F. Aryasetiawan and O. Gunnarsson},
   title = {The \textit{GW} method},
   journal = {Rep. Prog. Phys.},
   volume = {61},
   number = {},
   pages = {237},
   year = {1998},
   doi = {https://doi.org/10.1088/0034-4885/61/3/002}
}

@ARTICLE{vanSchilfgaarde2006PhysRevLett,
 author = {{van Schilfgaarde}, M. and Kotani, Takao and Faleev, S.},
 issn = {1079-7114},
 journal = {Phys. Rev. Lett.},
 month = {jun},
 number = {22},
 pages = {226402},
 publisher = {American Physical Society (APS)},
 title = {Quasiparticle Self-Consistent \textit{GW} Theory},
 doi = {http://dx.doi.org/10.1103/PhysRevLett.96.226402},
 volume = {96},
 year = {2006}
}

@ARTICLE{Bonacci2023npjComputMater,
 author = {Bonacci, Miki and Qiao, Junfeng and Spallanzani, Nicola and Marrazzo, Antimo and Pizzi, Giovanni and Molinari, Elisa and Varsano, Daniele and Ferretti, Andrea and Prezzi, Deborah},
 issn = {2057-3960},
 journal = {npj Comput. Mater.},
 month = {May},
 number = {1},
 publisher = {Springer Science and Business Media LLC},
 title = {Towards high-throughput many-body perturbation theory: efficient algorithms and automated workflows},
 doi = {http://dx.doi.org/10.1038/s41524-023-01027-2},
 volume = {9},
 year = {2023}
}

@ARTICLE{Louie2021NatMater,
 author = {Louie, Steven G. and Chan, Yang-Hao and da Jornada, Felipe H. and Li, Zhenglu and Qiu, Diana Y.},
 issn = {1476-4660},
 journal = {Nat. Mater.},
 month = {may},
 number = {6},
 pages = {728–735},
 publisher = {Springer Science and Business Media LLC},
 title = {Discovering and understanding materials through computation},
 doi = {http://dx.doi.org/10.1038/s41563-021-01015-1},
 volume = {20},
 year = {2021}
}

@ARTICLE{Blase2011PhysRevB,
 author = {Blase, X. and Attaccalite, C. and Olevano, V.},
 issn = {1550-235X},
 journal = {Phys. Rev. B},
 month = {Mar},
 number = {11},
 publisher = {American Physical Society (APS)},
 title = {First-principlesGWcalculations for fullerenes, porphyrins, phtalocyanine, and other molecules of interest for organic photovoltaic applications},
 doi = {http://dx.doi.org/10.1103/PhysRevB.83.115103},
 volume = {83},
 year = {2011},
 pages = 115103
}

@ARTICLE{yambo_2019,
  title = {Many-body perturbation theory calculations using the yambo code},
  author = {D. Sangalli and A. Ferretti and H. Miranda and C. Attaccalite and I. Marri and E. Cannuccia and P. Melo and M. Marsili and F. Paleari and A. Marrazzo and G. Prandini and P. Bonf{\`{a}} and M.~O. Atambo and F. Affinito and M. Palummo and A. Molina-S{\'{a}}nchez and C. Hogan and M. Gr{\" u}ning and D. Varsano and A. Marini},
  journal = {J. Phys.: Condens. Matter},
  volume = {31},
  issue = {32},
  pages = {325902},
  year = {2019},
  doi = {https://doi.org/10.1088/1361-648X/ab15d0},
}

@ARTICLE{paleari2025yambopy,
  title={YamboPy: a pre/post-processor tool for Yambo and QuantumEspresso written in Python.},
  author={Paleari, Fulvio and Molina-S{\'a}nchez, Alejandro and Nalabothula, Muralidhar and Reho, Riccardo and Bonacci, Miki and Castelo, Jos{\'e} and Cervantes-Villanueva, Jorge and Pionteck, Mike and Silvetti, Martino and Attaccalite, Claudio and others},
  year={2025},
  journal={Zenodo},
  doi={https://doi.org/10.5281/zenodo.15012962}
}

@ARTICLE{gonze2020abinit,
  title={The Abinit project: Impact, environment and recent developments},
  author={Gonze, Xavier and Amadon, Bernard and Antonius, Gabriel and Arnardi, Fr{\'e}d{\'e}ric and Baguet, Lucas and Beuken, Jean-Michel and Bieder, Jordan and Bottin, Fran{\c{c}}ois and Bouchet, Johann and Bousquet, Eric and others},
  journal={Computer Physics Communications},
  volume={248},
  pages={107042},
  year={2020},
  doi={gonze2020abinit},
  publisher={Elsevier}
}

@article{giannozzi2009JPCM_v2,
  title={QUANTUM ESPRESSO: a modular and open-source software project for quantum simulations of materials},
  author={Giannozzi, Paolo and Baroni, Stefano and Bonini, Nicola and Calandra, Matteo and Car, Roberto and Cavazzoni, Carlo and Ceresoli, Davide and Chiarotti, Guido L and Cococcioni, Matteo and Dabo, Ismaila and others},
  journal={Journal of physics: Condensed matter},
  volume={21},
  doi={https://doi.org/10.1088/0953-8984/21/39/395502},
  number={39},
  pages={395502},
  year={2009}
}

@article{giannozzi2017JPCM_v2,
  title={Advanced capabilities for materials modelling with Quantum ESPRESSO},
  author={Giannozzi, Paolo and Andreussi, Oliviero and Brumme, Thomas and Bunau, Oana and Buongiorno Nardelli, M and Calandra, Matteo and Car, Roberto and Cavazzoni, Carlo and Ceresoli, Davide and Cococcioni, Matteo and others},
  journal={Journal of physics: Condensed matter},
  volume={29},
  number={46},
  doi={https://doi.org/10.1088/1361-648X/aa8f79},
  pages={465901},
  year={2017},
  publisher={IOP Publishing}
}

@ARTICLE{Rozzi2006PRB,
  title = {Exact Coulomb cutoff technique for supercell calculations},
  author = {Rozzi, Carlo A. and Varsano, Daniele and Marini, Andrea and Gross, Eberhard K. U. and Rubio, Angel},
  journal = {Phys. Rev. B},
  volume = {73},
  issue = {20},
  pages = {205119},
  numpages = {13},
  year = {2006},
  doi = {https://doi.org/10.1103/PhysRevB.73.205119},
}

@ARTICLE{Cazzaniga2012PRB,
  title = {{$GW$} and beyond approaches to quasiparticle properties in metals},
  author = {Cazzaniga, Marco},
  journal = {Phys. Rev. B},
  volume = {86},
  issue = {3},
  pages = {035120},
  numpages = {9},
  year = {2012},
  doi = {https://doi.org/10.1103/PhysRevB.86.035120},
}

@ARTICLE{vonderLinden1988PRB,
  title = {Precise quasiparticle energies and Hartree-Fock bands of semiconductors and insulators},
  author = {{von der Linden}, Wolfgang and Horsch, Peter},
  journal = {Phys. Rev. B},
  volume = {37},
  issue = {14},
  pages = {8351--8362},
  numpages = {0},
  year = {1988},
  doi = {https://doi.org/10.1103/PhysRevB.37.8351},
}

@ARTICLE{Lee1994PRB,
  title = {First-principles study of the optical properties and the dielectric response of Al},
  author = {Lee, Keun-Ho and Chang, K. J.},
  journal = {Phys. Rev. B},
  volume = {49},
  issue = {4},
  pages = {2362--2367},
  numpages = {0},
  year = {1994},
  doi = {https://doi.org/10.1103/PhysRevB.49.2362},
}

@ARTICLE{Onida2002RMP,
  title={Electronic excitations: density-functional versus many-body Green’s-function approaches},
  author={Onida, Giovanni and Reining, Lucia and Rubio, Angel},
  journal={Rev. Mod. Phys.},
  volume={74},
  number={2},
  pages={601},
  year={2002},
  doi={https://doi.org/10.1103/RevModPhys.74.601}
}

@ARTICLE{Marzari2021NatureMat,
  title = {Electronic-structure methods for materials design},
  author = {Marzari, N. and Ferretti, A. and Wolverton, C.},
  journal = {Nature Materials},
  volume = {20},
  issue = {6},
  pages = {736–749},
  numpages = {15},
  year = {2021},
  doi = {https://doi.org/10.1038/s41563-021-01013-3},
}

@ARTICLE{Hedin1965PR,
  title = {New Method for Calculating the One-Particle Green's Function with Application to the Electron-Gas Problem},
  author = {Hedin, Lars},
  journal = {Phys. Rev.},
  volume = {139},
  issue = {3A},
  pages = {A796--A823},
  numpages = {0},
  year = {1965},
  doi = {https://doi.org/10.1103/PhysRev.139.A796},
}

@book{martin2016book,
  author={Martin,Richard M. and Reining,Lucia and Ceperley,David M.},
  title={Interacting Electrons},
  year=2016,
  publisher={Cambridge University Press},
  address={Cambridge},
  doi={https://doi.org/10.1017/CBO9781139050807}
}

@ARTICLE{Reining2018wcms,
author = {Reining, Lucia},
title = {The \textit{GW} approximation: content, successes and limitations},
journal = {WIREs Computational Molecular Science},
volume = {8},
number = {3},
pages = {e1344},
doi = {https://doi.org/10.1002/wcms.1344},
year = {2018}
}

@ARTICLE{Cazzaniga2010PRB,
  title = {Ab initio intraband contributions to the optical properties of metals},
  author = {Cazzaniga, Marco and Caramella, Lucia and Manini, Nicola and Onida, Giovanni},
  journal = {Phys. Rev. B},
  volume = {82},
  issue = {3},
  pages = {035104},
  numpages = {7},
  year = {2010},
  doi = {https://doi.org/10.1103/PhysRevB.82.035104},
}

@ARTICLE{Cazzaniga2008PRB,
  title = {Ab initio self-energy corrections in systems with metallic screening},
  author = {Cazzaniga, Marco and Manini, Nicola and Molinari, Luca Guido and Onida, Giovanni},
  journal = {Phys. Rev. B},
  volume = {77},
  issue = {3},
  pages = {035117},
  numpages = {6},
  year = {2008},
  doi = {https://doi.org/10.1103/PhysRevB.77.035117},
}

@ARTICLE{Krystyna_2020,
  title = {Impact of the Interband Transitions in Gold and Silver on the Dynamics of Propagating and Localized Surface Plasmons},
  author = {Krystyna Kolwas and Anastasiya Derkachova},
  journal = {Nanomaterials},
  volume = {10},
  issue = {7},
  pages = {1411},
  year = {2020},
  doi = {https://doi.org/10.3390/nano10071411}
}

@ARTICLE{Leon2021PRB,
  title = {Frequency dependence in \textit{GW} made simple using a multipole approximation},
  author = {Leon, Dario A. and Cardoso, Claudia and Chiarotti, Tommaso and Varsano, Daniele and Molinari, Elisa and Ferretti, Andrea},
  journal = {Phys. Rev. B},
  volume = {104},
  issue = {11},
  pages = {115157},
  numpages = {16},
  year = {2021},
  doi = {https://doi.org/10.1103/PhysRevB.104.115157},
}

@ARTICLE{Maksimov1988JPEMP,
	doi = {https://doi.org/10.1088/0305-4608/18/4/018},
	year = 1988,
	volume = {18},
	number = {4},
	pages = {833--849},
	author = {E~G Maksimov and I~I Mazin and S~N Rashkeev and Yu~A Uspenski},
	title = {First-principles calculations of the optical properties of metals},
	journal = {J. Phys. F: Metal Phys.},
}

@ARTICLE{Perdew1996,
  title = {Generalized Gradient Approximation Made Simple},
  author = {Perdew, John P. and Burke, Kieron and Ernzerhof, Matthias},
  journal = {Phys. Rev. Lett.},
  volume = {77},
  issue = {18},
  pages = {3865--3868},
  numpages = {0},
  year = {1996},
  doi = {https://doi.org/10.1103/PhysRevLett.77.3865}
}

@ARTICLE{Perdew-Zunger1981PRB,
author = {J P Perdew and A Zunger},
journal = {Phys. Rev. B},
title = {Self-interaction correction to density-functional approximations for many-electron systems},
affiliation = {TULANE UNIV,QUANTUM THEORY GRP,NEW ORLEANS,LA 70118},
number = {10},
pages = {5048--5079},
volume = {23},
year = {1981},
month = {Jan},
doi = {https://doi.org/10.1103/PhysRevB.23.5048}
}

@ARTICLE{Lyo1988PRL,
  title = {Quasiparticle band structure of Na and simple metals},
  author = {Lyo, In-Whan and Plummer, E.~W.},
  journal = {Phys. Rev. Lett.},
  volume = {60},
  issue = {15},
  pages = {1558--1561},
  numpages = {0},
  year = {1988},
  doi = {https://doi.org/10.1103/PhysRevLett.60.1558},
}

@ARTICLE{bartynski1986,
  title={Angle-resolved photoemission study of the surface and bulk electronic structure of Mg (0001) and Mg (112{\={}} 0)},
  author={Bartynski, RA and Gaylord, RH and Gustafsson, T and Plummer, EW},
  journal={Phys. Rev. B},
  volume={33},
  number={6},
  pages={3644},
  year={1986},
  publisher={APS},
  doi={https://doi.org/10.1103/PhysRevB.33.3644}
}

@ARTICLE{schiller2004,
  title={Electronic structure of Mg: From monolayers to bulk},
  author={Schiller, F and Heber, M and Servedio, VDP and Laubschat, C},
  journal={Phys. Rev. B},
  volume={70},
  number={12},
  pages={125106},
  year={2004},
  publisher={APS},
  doi={https://doi.org/10.1103/PhysRevB.70.125106}
}

@ARTICLE{Northrup1989PRB,
  title = {Quasiparticle excitation spectrum for nearly-free-electron metals},
  author = {Northrup, John E. and Hybertsen, Mark S. and Louie, Steven G.},
  journal = {Phys. Rev. B},
  volume = {39},
  issue = {12},
  pages = {8198--8208},
  numpages = {0},
  year = {1989},
  doi = {https://doi.org/10.1103/PhysRevB.39.8198}
}

@ARTICLE{Deslippe2012CPC,
title = {BerkeleyGW: A massively parallel computer package for the calculation of the quasiparticle and optical properties of materials and nanostructures},
journal = {Comput. Phys. Commun.},
volume = {183},
number = {6},
pages = {1269-1289},
year = {2012},
issn = {0010-4655},
doi = {https://doi.org/10.1016/j.cpc.2011.12.006},
author = {Jack Deslippe and Georgy Samsonidze and David~A. Strubbe and Manish Jain and Marvin~L. Cohen and Steven~G. Louie},
}

@ARTICLE{Orhan2019JPCM,
  title = {Plasmonic performance of AuxAgyCu1-x-y alloys from many-body perturbation theory},
  author = {Okan~K Orhan and David~D O'Regan },
  journal = {J. Phys.: Condens. Matter},
  volume = {31},
  issue = {31},
  pages = {315901},
  year = {2019},
  doi = {https://doi.org/10.1088/1361-648X/ab1c30},
}

@ARTICLE{DAmico2020PRB,
  title = {Intrinsic edge excitons in two-dimensional ${\mathrm{MoS}}_{2}$},
  author = {D'Amico, Pino and Gibertini, Marco and Prezzi, Deborah and Varsano, Daniele and Ferretti, Andrea and Marzari, Nicola and Molinari, Elisa},
  journal = {Phys. Rev. B},
  volume = {101},
  issue = {16},
  pages = {161410},
  numpages = {6},
  year = {2020},
  doi = {https://doi.org/10.1103/PhysRevB.101.161410},
}

@ARTICLE{Blochl1994PRB,
  title = {Improved tetrahedron method for Brillouin-zone integrations},
  author = {Bl\"ochl, Peter~E. and Jepsen, O. and Andersen, O.~K.},
  journal = {Phys. Rev. B},
  volume = {49},
  issue = {23},
  pages = {16223--16233},
  numpages = {0},
  year = {1994},
  doi = {https://doi.org/10.1103/PhysRevB.49.16223},
}

@ARTICLE{Methfessel1989PRB,
  title = {High-precision sampling for Brillouin-zone integration in metals},
  author = {Methfessel, M. and Paxton, A.~T.},
  journal = {Phys. Rev. B},
  volume = {40},
  issue = {6},
  pages = {3616--3621},
  numpages = {0},
  year = {1989},
  doi = {https://doi.org/10.1103/PhysRevB.40.3616},
}

@ARTICLE{Kohn1974PRB,
  title = {Wannier functions and self-consistent metal calculations},
  author = {Kohn, W.},
  journal = {Phys. Rev. B},
  volume = {10},
  issue = {2},
  pages = {382--383},
  numpages = {0},
  year = {1974},
  doi = {https://doi.org/10.1103/PhysRevB.10.382},
}

@ARTICLE{Sporkmann1994PRB,
  title = {Calculation of Wannier functions for fcc transition metals by Fourier transformation of Bloch functions},
  author = {Sporkmann, B. and Bross, H.},
  journal = {Phys. Rev. B},
  volume = {49},
  issue = {16},
  pages = {10869--10876},
  numpages = {0},
  year = {1994},
  doi = {https://doi.org/10.1103/PhysRevB.49.10869},
}

@ARTICLE{Marini2001PRB,
  title = {Plane-wave {DFT}-{LDA} calculation of the electronic structure and absorption spectrum of copper},
  author = {Marini, Andrea and Onida, Giovanni and Del Sole, Rodolfo},
  journal = {Phys. Rev. B},
  volume = {64},
  issue = {19},
  pages = {195125},
  numpages = {9},
  year = {2001},
  doi = {https://doi.org/10.1103/PhysRevB.64.195125},
}

@ARTICLE{Friedrichnano2022,
AUTHOR = {Friedrich, Christoph and Blügel, Stefan and Nabok, Dmitrii},
TITLE = {Quasiparticle Self-Consistent \textit{GW} Study of Simple Metals},
JOURNAL = {Nanomaterials},
VOLUME = {12},
YEAR = {2022},
NUMBER = {20},
pages ={3660},
ARTICLE-NUMBER = {3660},
doi = {https://www.mdpi.com/2079-4991/12/20/3660},
}

@ARTICLE{Strinati_1982,
  title = {Dynamical aspects of correlation corrections in a covalent crystal},
  author = {Strinati, G. and Mattausch, H. J. and Hanke, W.},
  journal = {Phys. Rev. B},
  volume = {25},
  issue = {4},
  pages = {2867--2888},
  numpages = {0},
  year = {1982},
  month = {Feb},
  publisher = {American Physical Society},
  doi = {https://doi.org/10.1103/PhysRevB.25.2867}
}

@ARTICLE{Prandini2019npjComputMater,
 author = {Prandini, Gianluca and Rignanese, Gian-Marco and Marzari, Nicola},
 issn = {2057-3960},
 journal = {npj Comput. Mater.},
 month = {Dec},
 number = {1},
 publisher = {Springer Science and Business Media LLC},
 title = {Photorealistic modelling of metals from first principles},
 doi = {http://dx.doi.org/10.1038/s41524-019-0266-0},
 volume = {5},
 pages = {129},
 year = {2019}
}

@ARTICLE{Prandini2019ComputPhysCommun,
 author = {Prandini, Gianluca and Galante, Mario and Marzari, Nicola and Umari, Paolo},
 issn = {0010-4655},
 journal = {Comput. Phys. Commun.},
 month = {Jul},
 pages = {106–119},
 publisher = {Elsevier BV},
 title = {SIMPLE code: Optical properties with optimal basis functions},
 doi = {http://dx.doi.org/10.1016/j.cpc.2019.02.016},
 volume = {240},
 year = {2019}
}

@book{Giuliani-Vignale2005book,
        author = {Giuliani, Gabriele and Vignale, Giovanni},
        title = {Quantum Theory of the Electron Liquid},
        publisher = {Cambridge University Press},
        year = {2005},
        doi={https://doi.org/10.1017/CBO9780511619915},
}

@book{Wooten1972book,
        author = {Frederick Wooten},
        title = {Optical properties of solids},
        publisher = {Academic Press},
        year = {1972},
        doi ={https://books.google.it/books?id=A_dHNRXFq28C},
}

@ARTICLE{monkhorst1976special,
  title={Special points for Brillouin-zone integrations},
  author={Monkhorst, Hendrik J and Pack, James D},
  journal={Phys. Rev. B},
  volume={13},
  number={12},
  pages={5188},
  year={1976},
  doi={https://doi.org/10.1103/PhysRevB.13.5188},
  publisher={APS}
}

@ARTICLE{Guandalini2023npjCM,
author={Guandalini, Alberto
and D'Amico, Pino
and Ferretti, Andrea
and Varsano, Daniele},
title={Efficient \textit{GW} calculations in two dimensional materials through a stochastic integration of the screened potential},
journal={npj Computational Materials},
year={2023},
month={Mar},
day={27},
volume={9},
number={1},
pages={44},
doi={https://doi.org/10.1038/s41524-023-00989-7}
}

@ARTICLE{GuandaliniLeon_2024,
  title = {Efficient {$GW$} calculations via interpolation of the screened interaction in momentum and frequency space: The case of graphene},
  author = {Guandalini, Alberto and Leon, Dario A. and D'Amico, Pino and Cardoso, Claudia and Ferretti, Andrea and Rontani, Massimo and Varsano, Daniele},
  journal = {Phys. Rev. B},
  volume = {109},
  issue = {7},
  pages = {075120},
  numpages = {15},
  year = {2024},
  month = {Feb},
  publisher = {American Physical Society},
  doi = {https://doi.org/10.1103/PhysRevB.109.075120}
}

@ARTICLE{Leon_2023,
Author = {Leon, Dario A. and Ferretti, Andrea and Varsano, Daniele and Molinari,
   Elisa and Cardoso, Claudia},
Title = {Efficient full frequency \textit{GW} for metals using a multipole approach for the dielectric screening},
Journal = {Phys. Rev. B},
Year = {2023},
Volume = {107},
Number = {15},
pages = {155130},
Month = {APR 17},
Publisher = {AMER PHYSICAL SOC},
Address = {ONE PHYSICS ELLIPSE, COLLEGE PK, MD 20740-3844 USA},
Type = {Article},
doi={https://doi.org/10.1103/PhysRevB.107.155130},
Article-Number = {155130},
}

@ARTICLE{Rasmussen_2016,
  title = {Efficient many-body calculations for two-dimensional materials using exact limits for the screened potential: Band gaps of {MoS}$_2$, $h$-{BN}, and phosphorene},
  author = {Rasmussen, Filip A. and Schmidt, Per S. and Winther, Kirsten T. and Thygesen, Kristian S.},
  journal = {Phys. Rev. B},
  volume = {94},
  issue = {15},
  pages = {155406},
  numpages = {9},
  year = {2016},
  month = {Oct},
  publisher = {American Physical Society},
  doi = {https://doi.org/10.1103/PhysRevB.94.155406}
}

@ARTICLE{daJornada_2017,
  title = {Nonuniform sampling schemes of the Brillouin zone for many-electron perturbation-theory calculations in reduced dimensionality},
  author = {da Jornada, Felipe H. and Qiu, Diana Y. and Louie, Steven G.},
  journal = {Phys. Rev. B},
  volume = {95},
  issue = {3},
  pages = {035109},
  numpages = {12},
  year = {2017},
  month = {Jan},
  publisher = {American Physical Society},
  doi = {https://doi.org/10.1103/PhysRevB.95.035109}
}

@ARTICLE{Xia_2020,
  title = {Combined subsampling and analytical integration for efficient large-scale \textit{GW} calculations for 2{D} systems},
  author = {Xia, Weiyi and Gao, Weiwei and Lopez-Candales, Gabriel and Wu, Yabei and Ren, Wei and Zhang, Wenqing and Zhang, Peihong},
  journal = {npj Comput. Mater.},
  volume = {6},
  issue = {1},
  pages = {118},
  year = {2020},
  doi = {https://doi.org/10.1038/s41524-020-00385-5}
}

@ARTICLE{qiu2016screening,
  title={Screening and many-body effects in two-dimensional crystals: Monolayer {MoS}$_2$},
  author={Qiu, Diana Y and Felipe, H and Louie, Steven G},
  journal={Phys. Rev. B},
  volume={93},
  number={23},
  pages={235435},
  year={2016},
  doi={https://doi.org/10.1103/PhysRevB.93.235435},
  publisher={APS}
}

@ARTICLE{Huser_2013,
  title = {How dielectric screening in two-dimensional crystals affects the convergence of excited-state calculations: Monolayer {MoS}$_2$},
  author = {H\"user, Falco and Olsen, Thomas and Thygesen, Kristian S.},
  journal = {Phys. Rev. B},
  volume = {88},
  issue = {24},
  pages = {245309},
  numpages = {9},
  year = {2013},
  month = {Dec},
  publisher = {American Physical Society},
  doi={https://doi.org/10.1103/PhysRevB.88.245309},
}

@ARTICLE{Marrazzo_2019,
author = {Marrazzo, Antimo and Gibertini, Marco and Campi, Davide and Mounet, Nicolas and Marzari, Nicola},
title = {Relative Abundance of {Z}2 Topological Order in Exfoliable Two-Dimensional Insulators},
journal = {Nano Letters},
volume = {19},
number = {12},
pages = {8431-8440},
year = {2019},
doi = {https://doi.org/10.1021/acs.nanolett.9b02689},
}

@ARTICLE{chiarotti2022unified,
  title={Unified {G}reen's function approach for spectral and thermodynamic properties from algorithmic inversion of dynamical potentials},
  author={Chiarotti, Tommaso and Marzari, Nicola and Ferretti, Andrea},
  journal={Phys. Rev. Res.},
  volume={4},
  number={1},
  pages={013242},
  year={2022},
  publisher={APS},
   doi ={http://dx.doi.org/10.1103/PhysRevResearch.4.013242},
}

@ARTICLE{AsgariMoS2,
  title = {Quasiparticle band-gap renormalization in doped monolayer {MoS}$_2$},
  author = {Faridi, Azadeh and Culcer, Dimitrie and Asgari, Reza},
  journal = {Phys. Rev. B},
  volume = {104},
  issue = {8},
  pages = {085432},
  numpages = {6},
  year = {2021},
  month = {Aug},
  publisher = {American Physical Society},
  doi = {https://doi.org/10.1103/PhysRevB.104.085432}
}

@ARTICLE{attaccalite2010doped,
  title={Doped graphene as tunable electron- phonon coupling material},
  author={Attaccalite, Claudio and Wirtz, Ludger and Lazzeri, Michele and Mauri, Francesco and Rubio, Angel},
  journal={Nano Letters},
  volume={10},
  number={4},
  pages={1172--1176},
  year={2010},
  publisher={ACS Publications},
  doi={https://doi.org/10.1021/nl9034626}
}

@ARTICLE{LiuMoS2,
  title = {Direct Determination of Band-Gap Renormalization in the Photoexcited Monolayer {MoS}$_2$},
  author = {Liu, Fang and Ziffer, Mark E. and Hansen, Kameron R. and Wang, Jue and Zhu, Xiaoyang},
  journal = {Phys. Rev. Lett.},
  volume = {122},
  issue = {24},
  pages = {246803},
  numpages = {6},
  year = {2019},
  month = {Jun},
  publisher = {American Physical Society},
  doi = {https://doi.org/10.1103/PhysRevLett.122.246803}
}

@ARTICLE{Spataru2004,
  title = {Excitonic Effects and Optical Spectra of Single-Walled Carbon Nanotubes},
  author = {Spataru, Catalin D. and Ismail-Beigi, Sohrab and Benedict, Lorin X. and Louie, Steven G.},
  journal = {Phys. Rev. Lett.},
  volume = {92},
  issue = {7},
  pages = {077402},
  numpages = {4},
  year = {2004},
  month = {Feb},
  publisher = {American Physical Society},
  doi = {https://doi.org/10.1103/PhysRevLett.92.077402}
}

@ARTICLE{Yang2007,
  title = {Quasiparticle Energies and Band Gaps in Graphene Nanoribbons},
  author = {Yang, Li and Park, Cheol-Hwan and Son, Young-Woo and Cohen, Marvin L. and Louie, Steven G.},
  journal = {Phys. Rev. Lett.},
  volume = {99},
  issue = {18},
  pages = {186801},
  numpages = {4},
  year = {2007},
  month = {Nov},
  publisher = {American Physical Society},
  doi = {https://doi.org/10.1103/PhysRevLett.99.186801}
}

@ARTICLE{spataru2010,
  title={Tunable Band Gaps and Excitons in Doped Semiconducting Carbon Nanotubes Made Possible by Acoustic Plasmons},
  author={Spataru, Catalin D and L{\'e}onard, Fran{\c{c}}ois},
  journal={Phys. Rev. Lett.},
  volume={104},
  number={17},
  pages={177402},
  year={2010},
  publisher={APS},
 doi = {https://journals.aps.org/prl/abstract/10.1103/PhysRevLett.104.177402}
}

@ARTICLE{Spataru2013,
title = {Quasiparticle and exciton renormalization effects in electrostatically doped semiconducting carbon nanotubes},
journal = {Chem. Phys.},
volume = {413},
pages = {81-88},
year = {2013},
note = {Photophysics of carbon nanotubes and nanotube composites},
issn = {0301-0104},
doi = {https://doi.org/10.1016/j.chemphys.2012.08.021},
author = {Catalin D. Spataru and François Léonard}
}

@ARTICLE{Liang2015,
  title = {Carrier Plasmon Induced Nonlinear Band Gap Renormalization in Two-Dimensional Semiconductors},
  author = {Liang, Yufeng and Yang, Li},
  journal = {Phys. Rev. Lett.},
  volume = {114},
  issue = {6},
  pages = {063001},
  numpages = {5},
  year = {2015},
  month = {Feb},
  publisher = {American Physical Society},
  doi = {https://doi.org/10.1103/PhysRevLett.114.063001}
}

@ARTICLE{Gao2017,
  title = {Renormalization of the quasiparticle band gap in doped two-dimensional materials from many-body calculations},
  author = {Gao, Shiyuan and Yang, Li},
  journal = {Phys. Rev. B},
  volume = {96},
  issue = {15},
  pages = {155410},
  numpages = {8},
  year = {2017},
  month = {Oct},
  publisher = {American Physical Society},
  doi = {https://doi.org/10.1103/PhysRevB.96.155410}
}

@ARTICLE{Champagne2023,
author = {Champagne, Aur{\'e}lie and Haber, Jonah B. and Pokawanvit, Supavit and Qiu, Diana Y. and Biswas, Souvik and Atwater, Harry A. and da Jornada, Felipe H. and Neaton, Jeffrey B.},
title = {Quasiparticle and Optical Properties of Carrier-Doped Monolayer {MoTe2} from First Principles},
journal = {Nano Letters},
volume = {23},
number = {10},
pages = {4274--4281},
year = {2023},
doi={https://doi.org/10.1021/acs.nanolett.3c00386},
}

@ARTICLE{braun1983plasmon,
  title={Plasmon excitation in {Al}-{Mg} surfaces},
  author={Braun, P and Arias, M and St{\"o}ri, H and Viehb{\"o}ck, FP},
  journal={Surf. Sci.},
  volume={126},
  number={1-3},
  pages={714--718},
  year={1983},
  publisher={Elsevier},
  doi ={https://doi.org/10.1016/0039-6028(83)90779-3}
}

@ARTICLE{steiner1978kll,
  title={The KLL Auger spectra of {Na} and {Mg} metal and their plasmon structure},
  author={Steiner, P and Reiter, FJ and H{\"o}chst, H and H{\"u}fner, S},
  journal={Physica Status Solidi (b)},
  volume={90},
  number={1},
  pages={45--51},
  year={1978},
  publisher={Wiley Online Library},
  doi={ https://doi.org/10.1002/pssb.2220900104}
}

@ARTICLE{smith1969photoemission,
  title={Photoemission studies of the alkali metals. I. Sodium and potassium},
  author={Smith, Neville V and Spicer, William E},
  journal={Phys. Rev.},
  volume={188},
  number={2},
  pages={593},
  year={1969},
  publisher={APS},
  doi={https://doi.org/10.1103/PhysRev.188.593}
}

@ARTICLE{Ugeda2014,
    author={Ugeda, M.M. and Bradley, A.J. and Shi,S.-F. and  daJornada, F.H. and  Zhang, Y. and  Qiu, D.Y. and  Ruan, W. and Mo,S.-K. and  Hussain, Z. and  Shen, Z.-X. and  Wang, F. and  Louie, S.G. and  Crommie, M.F.},
    title = {Giant bandgap renormalization and excitonic effects inamonolayer transition metal dichalcogenide semiconductor},
    journal = {Nat. Mater.},
    year = 2014,
    volume = 13,
    pages = {1091-1095},
    doi ={10.1038/nmat4061}
}

@ARTICLE{Mak2012,
    author = {Mak, K.F. and He, K. and Shan, J. and Heinz, T.F.},
    title = {Control of valley polarization inmonolayer {MoS}$_2$ byoptical helicity},
    journal = {Nat. Commun.},
    year = 2012,
    volume = 7,
    pages = {494-498},
    doi={https://doi.org/10.1038/nnano.2012.96}
}

@ARTICLE{Ross2013,
    author = {Ross, J. S. and Wu,S. and Yu,H. and Ghimire, N. J. and Jones, A. M. and  Aivazian, G. and Yan,J. and Mandrus, D. G. and Xiao, D. and Yao,W. and Xu,X.},
    title = {Electrical control of neutral and charged excitons in a monolayer semiconductor},
    journal = {Nat. Commun.},
    year = 2013,
    volume = 4,
    pages = 1474, 
    doi= {10.1038/ncomms2498}
}

@ARTICLE{Huser2013,
    author = {Hüser,F. and Olsen, T. and Thygesen, K. S.},
    title = {Quasiparticle \textit{GW} calculations for solids, molecules, and two-dimensional materials},
    journal = {Phys. Rev. B},
    year =  2013,
    volume = 87,
    pages = 235132,
    doi={https://doi.org/10.1103/PhysRevB.87.235132}
}

@ARTICLE{Adler1962PhysRev,
    author = "Adler, Stephen L.",
    issn = "0031-899X",
    journal = "Phys. Rev.",
    month = "April",
    number = "2",
    pages = "413-420",
    publisher = "American Physical Society (APS)",
    source = "Crossref",
    title = "Quantum Theory of the Dielectric Constant in Real Solids",
    doi = "https://doi.org/10.1103/physrev.126.413",
    volume = "126",
    year = "1962"
}

@ARTICLE{Wiser1963PhysRev,
    author = "Wiser, Nathan",
    issn = "0031-899X",
    journal = "Phys. Rev.",
    month = "January",
    number = "1",
    pages = "62-69",
    publisher = "American Physical Society (APS)",
    source = "Crossref",
    title = "Dielectric Constant with Local Field Effects Included",
    doi = "https://doi.org/10.1103/physrev.129.62",
    volume = "129",
    year = "1963"
}

@ARTICLE{hedin1999,
  title={On correlation effects in electron spectroscopies and the \textit{GW} approximation},
  author={Hedin, Lars},
  journal={J. Phys.: Condens. Matter},
  volume={11},
  number={42},
  pages={R489},
  year={1999},
  doi = {http://dx.doi.org/10.1088/0953-8984/11/42/201},
  publisher={IOP Publishing}
}

@ARTICLE{Lazzeri2006PhysRevLett,
 author = {Lazzeri, Michele and Mauri, Francesco},
 issn = {1079-7114},
 journal = {Phys. Rev. Lett.},
 month = {dec},
 number = {26},
 pages = {266407},
 publisher = {American Physical Society (APS)},
 title = {Nonadiabatic Kohn Anomaly in a Doped Graphene Monolayer},
 doi = {http://dx.doi.org/10.1103/PhysRevLett.97.266407},
 volume = {97},
 year = {2006}
}

@ARTICLE{albertoTB,
  title = {High- and low-energy many-body effects of graphene in a unified approach},
  author = {Guandalini, Alberto and Caldarelli, Giovanni and Macheda, Francesco and Mauri, Francesco},
  journal = {Phys. Rev. B},
  volume = {111},
  issue = {7},
  pages = {075118},
  numpages = {15},
  year = {2025},
  month = {Feb},
  publisher = {American Physical Society},
  doi = {https://doi.org/10.1103/PhysRevB.111.075118}
}

@ARTICLE{guandalini_2023,
author = {Guandalini, Alberto and Senga, Ryosuke and Lin, Yung-Chang and Suenaga, Kazu and Ferretti, Andrea and Varsano, Daniele and Recchia, Andrea and Barone, Paolo and Mauri, Francesco and Pichler, Thomas and Kramberger, Christian},
title = {Excitonic Effects in Energy-Loss Spectra of Freestanding Graphene},
journal = {Nano Letters},
volume = {23},
number = {24},
pages = {11835-11841},
year = {2023},
doi = { 
        https://doi.org/10.1021/acs.nanolett.3c03863
},
}

@ARTICLE{friedrich_2022,
	title = {Quasiparticle {Self}-{Consistent} \textit{GW} {Study} of {Simple} {Metals}},
	volume = {12},
	issn = {2079-4991},
	doi = {https://www.ncbi.nlm.nih.gov/pmc/articles/PMC9607527/},
	number = {20},
	urldate = {2025-08-06},
	journal = {Nanomaterials},
	author = {Friedrich, Christoph and Blügel, Stefan and Nabok, Dmitrii},
	month = oct,
	year = {2022},
	pmid = {36296848},
	pmcid = {PMC9607527},
	pages = {3660},
}

@ARTICLE{Lyo_88,
  title = {Quasiparticle band structure of Na and simple metals},
  author = {Lyo, In-Whan and Plummer, E. W.},
  journal = {Phys. Rev. Lett.},
  volume = {60},
  issue = {15},
  pages = {1558--1561},
  numpages = {0},
  year = {1988},
  month = {Apr},
  publisher = {American Physical Society},
  doi = {https://doi.org/10.1103/PhysRevLett.60.1558}
}

@ARTICLE{mandal_2022,
	title = {Electronic correlation in nearly free electron metals with beyond-{DFT} methods},
	volume = {8},
	copyright = {2022 The Author(s)},
	issn = {2057-3960},
	doi={https://doi.org/10.1038/s41524-022-00867-8},
	number = {1},
	urldate = {2025-08-06},
	journal = {npj Computational Materials},
	author = {Mandal, Subhasish and Haule, Kristjan and Rabe, Karin M. and Vanderbilt, David},
	month = aug,
	year = {2022},
	note = {Publisher: Nature Publishing Group},
	pages = {181},
}

@ARTICLE{stern1967,
  title={Polarizability of a two-dimensional electron gas},
  author={Stern, Frank},
  journal={Physical Review Letters},
  doi={https://doi.org/10.1103/PhysRevLett.18.546},
  volume={18},
  number={14},
  pages={546},
  year={1967},
  publisher={APS}
}

@ARTICLE{hwang2007dielectric,
  title={Dielectric function, screening, and plasmons in two-dimensional graphene},
  author={Hwang, EH and Das Sarma, S},
  journal={Phys. Rev. B},
  volume={75},
  doi={https://doi.org/10.1103/PhysRevB.77.081412},
  number={20},
  pages={205418},
  year={2007},
  publisher={APS}
}

@ARTICLE{da2020universal,
  title={Universal slow plasmons and giant field enhancement in atomically thin quasi-two-dimensional metals},
  author={da Jornada, Felipe H and Xian, Lede and Rubio, Angel and Louie, Steven G},
  journal={Nature communications},
  volume={11},
  number={1},
  pages={1013},
  year={2020},
  doi={https://doi.org/10.1038/s41467-020-14826-8},
  publisher={Nature Publishing Group UK London}
}

@ARTICLE{Jakovac2024,
  title = {Electron-plasmon scattering in doped graphene},
  author = {Jakovac, Josip and Despoja, Vito},
  journal = {Phys. Rev. B},
  volume = {110},
  issue = {19},
  pages = {195425},
  numpages = {13},
  year = {2024},
  month = {Nov},
  publisher = {American Physical Society},
  doi = {https://doi.org/10.1103/PhysRevB.110.195425}
}

@ARTICLE{DAlessio2025,
       author = {{D'Alessio}, Matteo and {Varsano}, Daniele and {Molinari}, Elisa and {Rontani}, Massimo},
        title = {Stabilization of sliding ferroelectricity through exciton condensation},
      journal = {arXiv e-prints},
         year = 2025,
        month = oct,
          eid = {arXiv:2510.01465},
        pages = {arXiv:2510.01465},
          doi={https://doi.org/10.48550/arXiv.2510.01465},
archivePrefix = {arXiv},
       eprint = {2510.01465}
}

@ARTICLE{Ataei2021,
	title = {Evidence of ideal excitonic insulator in bulk {MoS}$_{\textrm{2}}$ under pressure},
	volume = {118},
	issn = {0027-8424, 1091-6490},
	doi={https://doi.org/10.1073/pnas.2010110118},
	number = {13},
	journal = {Proceedings of the National Academy of Sciences},
	author = {Ataei, S. Samaneh and Varsano, Daniele and Molinari, Elisa and Rontani, Massimo},
	month = mar,
	year = {2021},
}

@ARTICLE{Bardeen1973,
  title = {Model for an Exciton Mechanism of Superconductivity},
  author = {Allender, David and Bray, James and Bardeen, John},
  journal = {Phys. Rev. B},
  volume = {7},
  issue = {3},
  pages = {1020--1029},
  numpages = {0},
  year = {1973},
  month = {Feb},
  publisher = {American Physical Society},
  doi = {https://doi.org/10.1103/PhysRevB.7.1020}
}

@ARTICLE{Cercellier2007,
  title = {Evidence for an Excitonic Insulator Phase in $1\text{T}\mathrm{\text{\ensuremath{-}}}\text{TiSe}_{2}$},
  author = {Cercellier, H. and Monney, C. and Clerc, F. and Battaglia, C. and Despont, L. and Garnier, M. G. and Beck, H. and Aebi, P. and Patthey, L. and Berger, H. and Forr\'o, L.},
  journal = {Phys. Rev. Lett.},
  volume = {99},
  issue = {14},
  pages = {146403},
  numpages = {4},
  year = {2007},
  month = {Oct},
  publisher = {American Physical Society},
  doi = {https://doi.org/10.1103/PhysRevLett.99.146403}
}

@ARTICLE{Crepel2022,
	title = {Spin-triplet superconductivity from excitonic effect in doped insulators},
	volume = {119},
	issn = {0027-8424, 1091-6490},
	doi={https://doi.org/10.1073/pnas.2117735119},
	number = {13},
	journal = {Proceedings of the National Academy of Sciences},
	author = {Crépel, Valentin and Fu, Liang},
	month = mar,
	year = {2022},
}

@ARTICLE{DiSalvo1976,
  title = {Electronic properties and superlattice formation in the semimetal ${\mathrm{TiSe}}_{2}$},
  author = {Di Salvo, F. J. and Moncton, D. E. and Waszczak, J. V.},
  journal = {Phys. Rev. B},
  volume = {14},
  issue = {10},
  pages = {4321--4328},
  numpages = {0},
  year = {1976},
  month = {Nov},
  publisher = {American Physical Society},
  doi = {https://doi.org/10.1103/PhysRevB.14.4321}
}

@ARTICLE{Gao2023,
	title = {Evidence of high-temperature exciton condensation in a two-dimensional semimetal},
	volume = {14},
	issn = {2041-1723},
	doi={https://doi.org/10.1038/s41467-023-36667-x},
	number = {1},
	journal = {Nature Communications},
	author = {Gao, Qiang and Chan, Yang-hao and Wang, Yuzhe and Zhang, Haotian and Jinxu, Pu and Cui, Shengtao and Yang, Yichen and Liu, Zhengtai and Shen, Dawei and Sun, Zhe and Jiang, Juan and Chiang, Tai C. and Chen, Peng},
	month = feb,
	year = {2023},
	pages = {994},
}

@ARTICLE{Gao2024,
	title = {Observation of possible excitonic charge density waves and metal–insulator transitions in atomically thin semimetals},
	volume = {20},
	issn = {1745-2481},
	doi={https://doi.org/10.1038/s41567-023-02349-0},
	number = {4},
	journal = {Nature Physics},
	author = {Gao, Qiang and Chan, Yang-hao and Jiao, Pengfei and Chen, Haiyang and Yin, Shuaishuai and Tangprapha, Kanjanaporn and Yang, Yichen and Li, Xiaolong and Liu, Zhengtai and Shen, Dawei and Jiang, Shengwei and Chen, Peng},
	month = apr,
	year = {2024},
	pages = {597--602},
}

@ARTICLE{Ginzburg1968,
	title = {The problem of high temperature superconductivity},
	volume = {9},
	issn = {0010-7514, 1366-5812},
	doi = {http://www.tandfonline.com/doi/abs/10.1080/00107516808220090},
	number = {4},
	journal = {Contemporary Physics},
	author = {Ginzburg, V. L.},
	month = jul,
	year = {1968},
	pages = {355--374},
}

@ARTICLE{Hossain2025,
	title = {Topological excitonic insulator with tunable momentum order},
	volume = {21},
	issn = {1745-2473, 1745-2481},
	doi={https://doi.org/10.1038/s41567-025-02917-6},
	number = {8},
	journal = {Nature Physics},
	author = {Hossain, Md Shafayat and Cheng, Zi-Jia and Jiang, Yu-Xiao and Cochran, Tyler A. and Zhang, Song-Bo and Wu, Huangyu and Liu, Xiaoxiong and Zheng, Xiquan and Cheng, Guangming and Kim, Byunghoon and Zhang, Qi and Litskevich, Maksim and Zhang, Junyi and Liu, Jinjin and Yin, Jia-Xin and Yang, Xian P. and Denlinger, Jonathan D. and Tallarida, Massimo and Dai, Ji and Vescovo, Elio and Rajapitamahuni, Anil and Yao, Nan and Keselman, Anna and Peng, Yingying and Yao, Yugui and Wang, Zhiwei and Balicas, Luis and Neupert, Titus and Hasan, M. Zahid},
	month = aug,
	year = {2025},
	pages = {1250--1259},
}

@ARTICLE{Ilani2016,
	title = {Electron attraction mediated by {Coulomb} repulsion},
	volume = {535},
	issn = {0028-0836, 1476-4687},
	doi={https://doi.org/10.1038/nature18639},
	number = {7612},
	journal = {Nature},
	author = {Hamo, A. and Benyamini, A. and Shapir, I. and Khivrich, I. and Waissman, J. and Kaasbjerg, K. and Oreg, Y. and Von Oppen, F. and Ilani, S.},
	month = jul,
	year = {2016},
	pages = {395--400},
}

@ARTICLE{Khveshchenko2001,
  title = {Ghost Excitonic Insulator Transition in Layered Graphite},
  author = {Khveshchenko, D. V.},
  journal = {Phys. Rev. Lett.},
  volume = {87},
  issue = {24},
  pages = {246802},
  numpages = {4},
  year = {2001},
  month = {Nov},
  publisher = {American Physical Society},
  doi = {https://doi.org/10.1103/PhysRevLett.87.246802}
}

@ARTICLE{Kogar2017,
	title = {Signatures of exciton condensation in a transition metal dichalcogenide},
	volume = {358},
	issn = {0036-8075, 1095-9203},
	doi={https://doi.org/10.1126/science.aam6432},
	number = {6368},
	journal = {Science},
	author = {Kogar, Anshul and Rak, Melinda S. and Vig, Sean and Husain, Ali A. and Flicker, Felix and Joe, Young Il and Venema, Luc and MacDougall, Greg J. and Chiang, Tai C. and Fradkin, Eduardo and van Wezel, Jasper and Abbamonte, Peter},
	month = dec,
	year = {2017},
	pages = {1314--1317},
}

@ARTICLE{Kono2017,
	title = {Zero-gap semiconductor to excitonic insulator transition in {Ta2NiSe5}},
	volume = {8},
	issn = {2041-1723},
	doi={https://doi.org/10.1038/ncomms14408},
	number = {1},
	journal = {Nature Communications},
	author = {Lu, Y. F. and Kono, H. and Larkin, T. I. and Rost, A. W. and Takayama, T. and Boris, A. V. and Keimer, B. and Takagi, H.},
	month = feb,
	year = {2017},
	pages = {14408},
}

@ARTICLE{Laussy2010,
  title = {Exciton-Polariton Mediated Superconductivity},
  author = {Laussy, Fabrice P. and Kavokin, Alexey V. and Shelykh, Ivan A.},
  journal = {Phys. Rev. Lett.},
  volume = {104},
  issue = {10},
  pages = {106402},
  numpages = {4},
  year = {2010},
  month = {Mar},
  publisher = {American Physical Society},
  doi = {https://doi.org/10.1103/PhysRevLett.104.106402}
}

@ARTICLE{Little1964,
  title = {Possibility of Synthesizing an Organic Superconductor},
  author = {Little, W. A.},
  journal = {Phys. Rev.},
  volume = {134},
  issue = {6A},
  pages = {A1416--A1424},
  numpages = {0},
  year = {1964},
  month = {Jun},
  publisher = {American Physical Society},
  doi = {https://doi.org/10.1103/PhysRev.134.A1416}
}

@ARTICLE{Littlewood1990,
	title = {Charge fluctuation mechanisms for high-temperature superconductivity},
	volume = {163},
	issn = {09214526},
	doi={https://doi.org/10.1016/0921-4526(90)90195-Z},
	number = {1-3},
	journal = {Physica B: Condensed Matter},
	author = {Littlewood, P.B.},
	month = apr,
	year = {1990},
	pages = {299--305},
}

@ARTICLE{Ma2021,
	title = {Strongly correlated excitonic insulator in atomic double layers},
	volume = {598},
	issn = {0028-0836, 1476-4687},
	doi={https://doi.org/10.1038/s41586-021-03947-9},
	number = {7882},
	journal = {Nature},
	author = {Ma, Liguo and Nguyen, Phuong X. and Wang, Zefang and Zeng, Yongxin and Watanabe, Kenji and Taniguchi, Takashi and MacDonald, Allan H. and Mak, Kin Fai and Shan, Jie},
	month = oct,
	year = {2021},
	pages = {585--589},
}

@ARTICLE{Portengen1996b,
  title = {Theory of electronic ferroelectricity},
  author = {Portengen, T. and \"Ostreich, Th. and Sham, L. J.},
  journal = {Phys. Rev. B},
  volume = {54},
  issue = {24},
  pages = {17452--17463},
  numpages = {0},
  year = {1996},
  month = {Dec},
  publisher = {American Physical Society},
  doi = {https://doi.org/10.1103/PhysRevB.54.17452}
}

@ARTICLE{Qi2025,
	title = {Perfect {Coulomb} drag and exciton transport in an excitonic insulator},
	volume = {388},
	issn = {0036-8075, 1095-9203},
	doi={https://doi.org/10.1126/science.adl1839},
	number = {6744},
	journal = {Science},
	author = {Qi, Ruishi and Joe, Andrew Y. and Zhang, Zuocheng and Xie, Jingxu and Feng, Qixin and Lu, Zheyu and Wang, Ziyu and Taniguchi, Takashi and Watanabe, Kenji and Tongay, Sefaattin and Wang, Feng},
	month = apr,
	year = {2025},
	pages = {278--283},
}

@ARTICLE{Sham1983,
  title = {Role of electron {C}oulomb interaction in superconductivity},
  author = {Rietschel, H. and Sham, L. J.},
  journal = {Phys. Rev. B},
  volume = {28},
  issue = {9},
  pages = {5100--5108},
  numpages = {0},
  year = {1983},
  month = {Nov},
  publisher = {American Physical Society},
  doi = {https://doi.org/10.1103/PhysRevB.28.5100}
}

@ARTICLE{Sham1984,
  title = {Superconductivity from nonphonon interactions},
  author = {Grabowski, M. and Sham, L. J.},
  journal = {Phys. Rev. B},
  volume = {29},
  issue = {11},
  pages = {6132--6142},
  numpages = {0},
  year = {1984},
  month = {Jun},
  publisher = {American Physical Society},
  doi = {https://doi.org/10.1103/PhysRevB.29.6132}
}

@ARTICLE{vanWezel2010,
  title = {Exciton-phonon-driven charge density wave in {T}i{S}e$_{2}$},
  author = {van Wezel, Jasper and Nahai-Williamson, Paul and Saxena, Siddarth S.},
  journal = {Phys. Rev. B},
  volume = {81},
  issue = {16},
  pages = {165109},
  numpages = {8},
  year = {2010},
  month = {Apr},
  publisher = {American Physical Society},
  doi = {https://doi.org/10.1103/PhysRevB.81.165109}
}

@ARTICLE{Varsano2020,
Author = {Varsano, Daniele and Palummo, Maurizia and Molinari, Elisa and Rontani,
   Massimo},
Title = {A monolayer transition-metal dichalcogenide as a topological excitonic
   insulator},
Journal = {Nature Nanotechnology},
Year = {2020},
Volume = {15},
Number = {5},
Pages = {367},
Month = {MAY},
doi={https://doi.org/10.1038/s41565-020-0650-4},
ISSN = {1748-3387},
EISSN = {1748-3395}
}

@ARTICLE{Wu2024,
Author = {Wu, Sanfeng and Schoop, Leslie M. and Sodemann, Inti and Moessner,
   Roderich and Cava, Robert J. and Ong, N. P.},
Title = {Charge-neutral electronic excitations in quantum insulators},
Journal = {Nature},
Year = {2024},
Volume = {635},
Number = {8038},
Pages = {301-310},
Month = {NOV 14},
doi={https://doi.org/10.1038/s41586-024-08091-8},
ISSN = {0028-0836},
EISSN = {1476-4687}
}

@ARTICLE{zheng_2023,
  title={{Electronic ratchet effect in a moir\'e system: signatures of excitonic ferroelectricity}},
  author={Zhiren Zheng and Xueqiao Wang and Ziyan Zhu and Stephen Carr and Trithep Devakul and Sergio de la Barrera and Nisarga Paul and Zumeng Huang and Anyuan Gao and Yang Zhang and Damien Bérubé and Kathryn Natasha Evancho and Kenji Watanabe and Takashi Taniguchi and Liang Fu and Yao Wang and Su-Yang Xu and Efthimios Kaxiras and Pablo Jarillo-Herrero and Qiong Ma},
  journal={arXiv:2306.03922},
  year={2023},
  doi ={https://doi.org/10.48550/arXiv.2306.03922},
}

@ARTICLE{fei_ferroelectric_2018,
	title = {Ferroelectric switching of a two-dimensional metal},
	volume = {560},
	issn = {0028-0836, 1476-4687},
	doi={https://doi.org/10.1038/s41586-018-0336-3},
	number = {7718},
	journal = {Nature},
	author = {Fei, Zaiyao and Zhao, Wenjin and Palomaki, Tauno A. and Sun, Bosong and Miller, Moira K. and Zhao, Zhiying and Yan, Jiaqiang and Xu, Xiaodong and Cobden, David H.},
	month = aug,
	year = {2018},
	pages = {336--339},
}

@ARTICLE{jia_evidence_2022,
	title = {Evidence for a monolayer excitonic insulator},
	volume = {18},
	issn = {1745-2473, 1745-2481},
	doi={https://doi.org/10.1038/s41567-021-01422-w},
	number = {1},
	journal = {Nature Physics},
	author = {Jia, Yanyu and Wang, Pengjie and Chiu, Cheng-Li and Song, Zhida and Yu, Guo and Jäck, Berthold and Lei, Shiming and Klemenz, Sebastian and Cevallos, F. Alexandre and Onyszczak, Michael and Fishchenko, Nadezhda and Liu, Xiaomeng and Farahi, Gelareh and Xie, Fang and Xu, Yuanfeng and Watanabe, Kenji and Taniguchi, Takashi and Bernevig, B. Andrei and Cava, Robert J. and Schoop, Leslie M. and Yazdani, Ali and Wu, Sanfeng},
	month = jan,
	year = {2022},
	pages = {87--93},
}

@ARTICLE{jindal_coupled_2023,
	title = {Coupled ferroelectricity and superconductivity in bilayer {Td}-{MoTe2}},
	volume = {613},
	copyright = {2023 The Author(s), under exclusive licence to Springer Nature Limited},
	issn = {1476-4687},
	doi={https://doi.org/10.1038/s41586-022-05521-3},
	number = {7942},
	journal = {Nature},
	author = {Jindal, Apoorv and Saha, Amartyajyoti and Li, Zizhong and Taniguchi, Takashi and Watanabe, Kenji and Hone, James C. and Birol, Turan and Fernandes, Rafael M. and Dean, Cory R. and Pasupathy, Abhay N. and Rhodes, Daniel A.},
	month = jan,
	year = {2023},
	pages = {48--52},
}

@ARTICLE{sun_evidence_2022,
	title = {Evidence for equilibrium exciton condensation in monolayer {WTe$_2$}},
	volume = {18},
	issn = {1745-2473, 1745-2481},
	doi={https://doi.org/10.1038/s41567-021-01427-5},
	number = {1},
	journal = {Nature Physics},
	author = {Sun, Bosong and Zhao, Wenjin and Palomaki, Tauno and Fei, Zaiyao and Runburg, Elliott and Malinowski, Paul and Huang, Xiong and Cenker, John and Cui, Yong-Tao and Chu, Jiun-Haw and Xu, Xiaodong and Ataei, S. Samaneh and Varsano, Daniele and Palummo, Maurizia and Molinari, Elisa and Rontani, Massimo and Cobden, David H.},
	month = jan,
	year = {2022},
	pages = {94--99},
}

\end{document}